\documentclass[twocol]{ametsocV6.1}
\usepackage{xcolor}
\usepackage{siunitx}
\usepackage{pgfplots}
\usepackage{pgfcalendar}
\usepackage{subcaption}
\usepackage{multirow}
\usepackage{textcomp}
\usepackage[nameinlink]{cleveref}
\pgfplotsset{width=10cm, compat=1.18}
\usepgfplotslibrary{groupplots}
\usepgfplotslibrary{dateplot}
\usepgfplotslibrary{fillbetween}
\usetikzlibrary{patterns}

\definecolor{jazzberry_jam}{RGB}{159, 1, 98}
\definecolor{jazzberry_creepers}{RGB}{0, 159, 129}
\definecolor{barbie_pink}{RGB}{255, 90, 175}
\definecolor{aquamarine}{RGB}{0, 252, 207}
\definecolor{french_violet}{RGB}{132, 0, 205}
\definecolor{dodger_blue}{RGB}{0, 141, 249}
\definecolor{capri}{RGB}{0, 194, 249}
\definecolor{plum}{RGB}{255, 178, 253}
\definecolor{carmine}{RGB}{164, 1, 34}
\definecolor{alizarin_crimson}{RGB}{226, 1, 52}
\definecolor{outrageous_orange}{RGB}{255, 110, 58}
\definecolor{bright_spark}{RGB}{255, 195, 59}

\usepackage{pgfplots}
\pgfplotsset{compat=newest}
\usepgfplotslibrary{external}
\usepackage{pdfpages}

\title{Global Forecasting of Tropical Cyclone Intensity Using Neural Weather Models}

\authors{Milton Gomez\aff{a,b}\correspondingauthor{Milton Gomez, milton.gomez@unil.ch}, Louis Poulain-$\,$-Auzeau\aff{a,b}, Alexis Berne\aff{c}, Tom Beucler\aff{a,b}}

\affiliation{\aff{a}{Faculty of Geosciences and Environment, University of Lausanne, Lausanne, VD, Switzerland}\\  
\aff{b}{Expertise Center for Climate Extremes, University of Lausanne, Lausanne, VD, Switzerland}\\  
\aff{c}{Environmental Remote Sensing Laboratory, EPFL, Lausanne, VD, Switzerland}}

\abstract{
Numerical Weather Prediction (NWP) models that integrate coupled physical equations forward in time are the traditional tools for simulating atmospheric processes and forecasting weather. With recent advancements in deep learning, AI-based Weather Prediction models that rely on neural network architectures--Neural Weather Models (NeWMs)--have emerged as competent medium-range NWP emulators, with performances that compare favorably to state-of-the-art NWP models. However, they are commonly trained on reanalyses with limited spatial resolution (e.g., 0.25$^{\circ}$ horizontal grid spacing), which smooths out key features of weather systems. For example, tropical cyclones (TCs)—among the most impactful weather events due to their devastating effects on human activities—are challenging to forecast, as extrema 
are smoothed in deterministic forecasts at 0.25$^{\circ}$ resolution. To address this, we use our best observational estimates of wind gusts and minimum sea level pressure to train a hierarchy of post-processing models on NeWM outputs. Applied to Pangu-Weather and FourCastNet v2, the post-processing models produce accurate and reliable forecasts of TC intensity up to five days ahead. Our post-processing algorithm is tracking-independent, preventing full misses, and we demonstrate that even linear models extract predictive information from NeWM outputs beyond what is encoded in their initial conditions. While spatial masking improves probabilistic forecast consistency, we do not find clear advantages of convolutional architectures over simple multilayer perceptrons for our NeWM post-processing purposes. Overall, by combining the efficiency of NeWMs with a lightweight, tracking-independent post-processing framework, our approach improves the accessibility of global TC intensity forecasts, marking a step toward their democratization.
}

\begin{document}

\maketitle

%
%
%
\statement
Forecasting tropical cyclone intensity via purely data-driven methods is limited to short lead times without large-scale atmospheric context. AI global weather models predict this context but at resolutions too coarse to resolve extremes such as those associated with tropical cyclones. We show that post-processing these global models with neural networks yields well-calibrated probabilistic forecasts of tropical cyclone intensity up to five days ahead. This work furthers end-to-end, fully data-driven forecasting of weather extremes.

\section{Introduction}

Over the past five decades, there have been significant advances in tropical cyclone (TC) track prediction \citep{demaria_is_2014}, largely due to increased computational power and improved remote sensing of the tropical atmosphere. However, traditional numerical weather prediction (NWP) models continue to struggle with forecasting TC intensity \citep{emanuel2016predictability}, particularly rapid intensification, which is defined as a sharp increase in maximum sustained winds \citep{elsberry2007accuracy}. These limitations partly stem from errors in the initial conditions, boundary layer physics, and the predicted TC environment.

Recent advances in machine learning are beginning to address some of these challenges, particularly in forecasting the TC environment. Unlike traditional NWP models, the current generation of global data-driven models is trained primarily on ERA5 \citep{hersbach_era5_2020}, the fifth-generation reanalysis from the European Centre for Medium-Range Weather Forecasts (ECMWF). ERA5 provides continuous data from 1940 to the present for hundreds of variables, resolved hourly at $0.25^\circ$ horizontal spatial resolution and with over 30 pressure levels vertically. Following earlier progress in nowcasting (i.e., forecasts for the next few hours), recent years have witnessed an AI revolution in medium-range forecasting \citep{bouallegue2024,beucler2024next}. This shift has been driven by the development of several purely data-driven deep learning models \citep{lam_learning_2023, bi_accurate_2023, pathak_fourcastnet_2022, bonev_spherical_2023}, often referred to as AI Weather Prediction (AIWP) models and which achieve deterministic errors comparable to those of the best NWP models \citep{rasp_weatherbench_2023}. Since the AIWP models investigated in this study rely on neural network-based architectures, we will refer to these as Neural Weather Models (NeWMs) to distinguish them from other AIWP models that may not rely on deep learning. Some NeWMs now also produce skillful probabilistic forecasts, outperforming traditional ensemble systems in predicting extreme weather, tropical cyclone tracks, and wind power production \citep{gencast24}. A major advantage of NeWMs is their ability to generate forecasts within minutes on relatively inexpensive hardware, in contrast to the thousands of CPU core hours often required by traditional NWP models \citep{Michalakes2020}.

Evaluating NeWMs, however, remains complex due to the multifaceted nature of weather forecasting and the diverse interests of stakeholders. Community platforms such as WeatherBench \citep{rasp_weatherbench_2020} have evolved to support more flexible evaluation metrics and now include probabilistic scoring \citep{rasp_weatherbench_2023}, reflecting a preference for a range of predictions rather than single outcomes.

While NeWMs are widely evaluated through such platforms, holistic evaluation on specific tasks remains limited. This is especially relevant because machine learning models are often trained to minimize mean squared error (MSE), which tends to improve performance of bulk properties at the expense of rare or small-scale meteorological events. Community efforts are beginning to assess NeWMs for TC track and intensity forecasting \citep{demaria2024evaluation}. Early post-processing efforts in generative settings show promise for downscaling applications \citep{jing2024tc,lockwood2024}, but such methods are restricted to basins with high-resolution TC wind field analyses (e.g., HWind for hurricanes, \cite{powell1998hrd}), and cannot currently be generalized globally. Simpler alternatives that learn a direct mapping from ERA5 to observed TC intensity bypass this bottleneck but suffer from domain shift when applied to NeWM-generated fields \citep{jing2024tc}. Other studies have shown that NeWM outputs can be valuable for assessing severe convective outlooks \citep{feldmann2024lightning}. Yet, to our knowledge, no study has tested whether global NeWMs can be effectively post-processed to yield improved global TC intensity forecasts. This gap is problematic as machine learning-based TC intensity predictions rarely exceed 24-48 hour lead times \citep{Meng_2023,griffin2024predicting,Gupta2025} without information from global weather forecasts. While post-processing has proven effective for station-scale wind and temperature forecasts \citep{bremnes2023evaluation}, data-driven models often underfit extremes due to their loss functions \citep{xu2024extremecast}. Their capacity to reliably forecast extremes hence remains an open question \citep{bülte2024uncertainty, olivetti2024data}.

In the context of a warming climate, this challenge becomes increasingly urgent. Although the impact of climate change on TC frequency remains an open question~\citep[e.g.,][]{sobel2021tropical}, TC intensities are projected to rise \citep{kang2015intensity, elsner2008intensity, sobel2016humanintensity} as are the frequency of rapid intensification events \citep{grondin2024climatology, bhatia2022potential}. In a 2$^{\circ}$C warming scenario, TC intensity is expected to increase by about 5\%, while the median projected change in the frequency of category 4–5 storms is a 13\% increase~\citep{knutson_tropical_2020}. In this context, accurate prediction of extreme TCs and their uncertainty is key for climate adaptation, and likely requires post-processing.

\begin{figure*}[h]
    \centering
    \includegraphics[width=\textwidth]{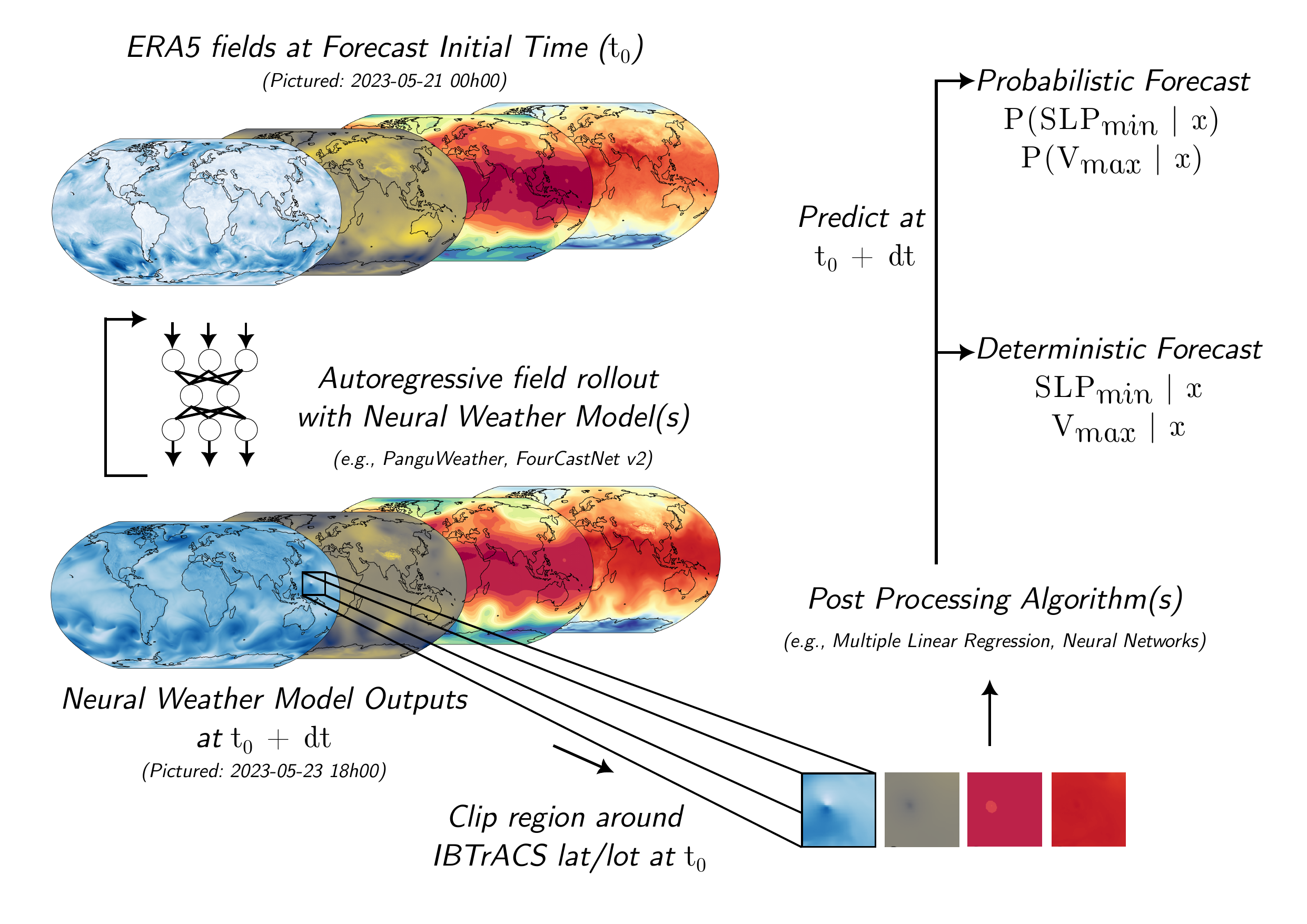}
    \caption{We propose post-processing NeWM forecasts to improve the prediction of TC intensity. Our postprocessing pipeline includes three steps: first, we generate forecast fields using neural weather models. Second, we clip the global fields produced by the models to a bounded region, using the reported location from IBTrACS at the initial time of forecast. Finally, we try to match observation targets using a deterministic or a probabilistic post-processing model. }
    \label{fig:pipeline}
\end{figure*}

In this study, rather than only exploring the direct abilities of AI-based weather models to predict TC intensity, we analyze their ability to provide a $0.25^\circ$-scale environment that supports improved TC intensity forecasts through post-processing, as illustrated in Figure.~\ref{fig:pipeline}. Our evaluation approach deliberately avoids relying on storm tracking within NeWMs, since storm detection in ERA5-like fields is sensitive to the choice of tracking algorithm \citep{bourdin22}, which can lead to full misses and false alarms.

Two main factors motivate our focus on post-processing global NeWMs rather than traditional NWP forecasts, as done in \cite{kieu2025nwp}. First, from an accessibility standpoint, post-processing AI model outputs with deep learning enables an end-to-end TC forecasting system that can run on a modern laptop. Second, and more fundamentally, we show that global AI weather models capture nontrivial patterns beyond what can be extracted from initial conditions alone. While our framework is tailored to NeWMs, it remains compatible with standard NWP output, provided the necessary meteorological fields are available. 

\section{Data}\label{sec:Dataset}

The post-processing framework developed in our study requires handling meteorological reanalysis, AI model, and observational data. In this section, we present the different data sources and how they are used in the framework.

\subsection{Observational Target: IBTrACS}\label{subsec:ibtracs}

The end-goal of the post-processing pipeline is to produce a better prediction of TC intensity. However, as there is a variation in the definitions of TC intensity across the various meteorological agencies in charge of tracking TCs~\citep{Neumann2017tropical, schreck2014impact}, in our study we define the intensity as the \mbox{1-minute} maximum sustained wind speed at 10 meters, $V_{max}$ (in knots$\approx0.51\text{m}\ \text{s}^{-1}$) and the minimum sea-level pressure, $P_{min}$ (in hPa). We have selected this definition to align with the values reported by United States Agencies in IBTrACS ~\citep{knapp_international_2010, knapp2018ibtracs}, as US agencies provide intensity values for all storms they detect across global basins. These IBTrACs ($V_{max},P_{min}$) pairs establish a regression target for the post-processing algorithm (the "ground truth"), noting that trustworthy observations are needed to ensure the algorithm does not learn from unrealistic samples. This presents a challenge in the database given the increase in uncertainty for older records, especially for those dating before reliable satellite imagery. As a first step, we limit the analysis to the most recent ten years: 2013 to 2023. 

To construct the ground truth, we begin by filtering the IBTrACS dataset to retain the 2013-2023 period, whereupon we only keep timestamps corresponding to 00:00, 06:00, 12:00, and 18:00. Our timestamp selection is done because IBTrACS presents a 3-hourly record, which relies on interpolating intensity outside of the 6 hourly reports when 3-hourly records are unavailable. Further, our selection removes any extraordinary reports (e.g., those covering intensity at landfall).  We then select a series of lead times ($\tau$) that we were interested in predicting for at each timestamp ($\tau \in \left\{6,\ 12,\ 18,\ 24,\ 48,\ 72,\ 96,\ 120,\ 144,\ 168\right\}\mathrm{h}$)--noting that 168h correspongs to 7 days. The lead times were selected to provide thorough examination of short lead times (where simple baselines such as persistence show significant skill), and a daily evaluation of intensity that is ambitious (extending beyond the traditional 5-day predictions reported by the National Hurricane Center and ECWMF) but within the predictability window of medium-range weather forecasting.
Then, for each timestamp $t$ we verify that a corresponding truth value exists at $t+\tau$, and remove instances where a truth value is not available (e.g., when there is a series gap due to de-intensification or when gaps exist due to storms undergoing extratropical transition). We further note the limitation that IBTrACS reports intensity value in stepped form with a resolution of 5 kt for $V_{max}$ and 5 hPa for $P_{min}$, inducing possible quantization effects.

\subsection{Meteorological Reanalysis: ERA5}

ERA5~\citep{hersbach_era5_2020} is a global reanalysis dataset developed by the ECMWF and is, up to this day, considered the best estimate of the atmospheric states in past decades~\citep{guo_investigation_2021}. The NeWMs used in this study were trained on ERA5, using it both to provide initial conditions and as the target for autoregressive forecasting. These models can be fine-tuned for operational use with ECMWF analysis data~\citep{rasp_weatherbench_2023}. In our study, we rely on the implementation of the NeWMs provided by the ECMWF in its \mbox{AI-Models} repository~\citep{AImodels}. 

We thus rely on the ERA5 single-level and vertical-level variables to provide each NeWM with the initial conditions each model requires. 
While ERA5 offers a consistent estimation of the atmospheric state, it does not fully capture the intensity of TCs as recorded in best-track datasets like IBTrACS. We thus discuss this disparity in the following subsection. 

\subsection{Comparing ERA5 and IBTrACS}

As noted in the previous subsections, the TC intensity values reported in IBTrACS and ERA5 differ in definition, which complicates direct comparison.

Just as there is variability in TC intensity definitions among the reporting agencies within IBTrACS, ERA5 does not offer wind and pressure fields that directly match these definitions. ERA5 provides instantaneous 10-meter wind fields on a $0.25^\circ$ grid. Meteorological reanalyses tend to underestimate TC intensities\citep{hodges2017well}, particularly for $V_{max}$, beyond what coarse resolution alone would yield \citep{schenkel2012}. In contrast, $P_{min}$ fields are somewhat better represented in ERA5 \citep{dulac_assessing_2024}. 

Discrepancies also arise from the reporting format: IBTrACS provides $V_{max}$ and $P_{min}$ in 5-knot and 5-hPa increments, while ERA5 values are continuous. Moreover, ERA5 fields represent grid-cell averages, whereas IBTrACS values are point estimates. These differences, both in spatial resolution and reporting granularity, make direct mapping between the two datasets inherently difficult. 

Finally, ERA5 is a reanalysis product that results from assimilating a wide range of observations into a numerical weather model. Its accuracy depends on both the quality of the assimilated data (e.g., satellite observations) and the subgrid-scale parameterizations used in the IFS model. As a result, the number and characteristics of TCs found using tracking algorithms applied to ERA5 do not always match observational records \citep{bourdin22}.

Given these discrepancies, NeWMs trained to emulate ERA5 cannot, at present, directly produce TC intensity estimates consistent with IBTrACS. This necessitates post-processing. 

\section{Global Medium-Range Neural Weather Models}

Deep learning has accelerated the development of medium-range, data-driven weather forecasting models. For example, Graphcast, FourCastNet and PanguWeather --amongst others-- now provide deterministic forecasts that are competitive with IFS HRES\footnote{Integrated Forecasting System High Resolution Ensemble System}, as measured by the RMSE (Root Mean Squared Error). While these models are still being evaluated in a deterministic way, the community is moving towards probabilistic predictions and evaluations~\citep{chapman_probabilistic_2022, garg_weatherbench_2022, rasp_weatherbench_2023, gencast24}. 
\par
In the following subsections, we provide a brief overview of the models used in this study, noting that the models were chosen for easy access to them on public repositories at the time of the start of this study (February 2024). Since then, other NeWMs have been developed~\citep{ chen_fuxi_2023, kochkov_neural_2024, lang2024aifs} and could be readily post-processed using our framework. 

\subsection{Neural Weather Model Characteristics}

\subsubsection{PanguWeather}

Developed by Huawei in 2022~\citep{bi_accurate_2023}, utilizes the transformer architecture introduced by~\cite{vaswani_attention_2017}. Originally designed for word translation, transformers use three key concepts: embedding, which transforms input data into vectors; self-attention layers, which weigh the importance of different parts of an input sequence to capture long-range dependencies; and positional encoding, which provides information about the order of data points in a sequence. PanguWeather processes a combination of surface and atmospheric fields by performing embedding operations on each type of field separately before concatenating the results. 

\subsubsection{FourCastNet v2}

Developed by Nvidia in 2023~\citep{bonev_spherical_2023}, FourCastNet version 2 relies on the concept of neural operators, introduced by~\cite{chen1995universal, lu2019deeponet}, and uses spherical neural operators to address shortcomings such as singularity near the Poles in the original architecture ~\citep{pathak_fourcastnet_2022}. Neural operators are designed to map between infinite-dimensional function spaces, which may allow the information learned by the neural network to be used across different resolutions and meshes~\citep{kochkov_neural_2024}. While in practice there are challenges associated with generalizing across unseen resolutions and meshes~\citep[e.g.,][]{liu2023domain}, neural operator based architectures can be made computationally efficient with good performance across resolutions. As we use the AI-Models repository, we specifically rely on the model released under the name ``FourCastNetv2-small''.

\subsection{Generating Global Weather Forecasts}

NeWMs have generally been configured to output prescribed variable fields spanning the whole globe at a given horizontal and vertical resolution. As with NWP models, the NeWMs used in this study require a set of starting conditions before being rolled out auto-regressively. To this end, we extract the ERA5 fields required to run each NeWM at each of the six-hourly timestamps established in subsection~\ref{sec:Dataset}\ref{subsec:ibtracs}. In order to run the models, we rely on the implementations made available by the ECMWF~\citep{AImodels} with a notable change in that we modify the input pipeline to work off of local storage rather than relying on on-the-fly downloading from the Copernicus Climate Data Store. Because of the autoregressive nature of the NeWMs, we generate data not only for the desired lead times, but also for any previous lead time needed to generate the fields at the desired lead times. For example, given that FourCastNetv2 is designed to provide 3-hourly forecasts we generate a total of 28 predictions, for which we only keep the fields associated with the 10 lead times we use for our study (i.e., $\tau \in \left\{6, 12, 18, 24, 48, 72, 96, 120, 144, 168\right\}\mathrm{h}$). With regards to the meteorological variables selected in our framework, we choose the meridional and zonal components of the 10-m wind ($u_{10m}$, $v_{10m}$), the mean sea level pressure ($P_0$), the 500 hPa geopotential height ($z_{500}$), and the 850 hPa temperature ($T_{850}$). The meridional and zonal components of wind, as well as the mean sea level pressure, were selected as they are natural analogues for $V_{max}$ and $P_{min}$. Meanwhile, $z_{500}$ and $T_{850}$ are standard benchmarking variables for NeWMs and are hence generally well predicted~\citep{rasp_weatherbench_2023}. To facilitate interpretation and normalization, the 10-m wind component fields are transformed into the magnitude of the 10-m horizontal wind ($|\mathbf{V}_{10m}|$) and the orientation of the 10-m wind ($\vartheta_{10m}$)

\subsection{Bypassing Tracking Using Initial Storm Conditions}\label{subsub:FieldPrep}
As previously mentioned, NeWMs are generally configured to produce a global forecast of the predicted fields. Traditionally, a tracking algorithm (e.g., the ``TempestExtremes'' tracking algorithm described in~\cite{ullrich2017tempestextremes}) is used to detect storm tracks in NWP outputs. The detected tracks are matched to the closest storm that exists in the observed record, and the information about the predicted position is used to determine the predicted intensity using a prescribed algorithm. While~\cite{demaria_is_2014} describe using the US Global Forecasting System--GFS--~\citep{GFSDoc} and finding nearly identical tracks when using a simplified tracker on (GFSS) and an operational tracker (GFSO), \cite{bourdin22} report appreciable differences between the behaviors of four different trackers (the tracker in~\citep{ullrich2017tempestextremes}, a tracker based on the Okubo-Weiss Parameter~\citep{tory2013importance}, the TRACK method described by~\citep{strachan2013investigating, hodges2017well}, and the tracker from the French National Center for Meteorological Research--CNRM~\citep{chauvin2006response}) when applied to ERA5. Given that the NeWMs used in our study are trained on ERA5 data, we decided to control for the effect of the tracking algorithm by \textit{omitting} the use of a tracker. Instead, we define storm-centered input domains based on the statistical behavior of storms in the training set, as detailed below:

\subsubsection{Initial Conditions} We assume that the tropical system of interest is known to exist, and use its initial intensity $\left(V_{max},P_{min}\right)$ and location at the \textit{time of forecast}(i.e., the time for which we have an initial state from which to make a forecast). We note that this time is often referred to as the \textit{initial time}, $t_0$--which is not to be confused with the time of cyclogenesis which is when the tropical system is considered to become a tropical cyclone.

\subsubsection{Domain Extent} 
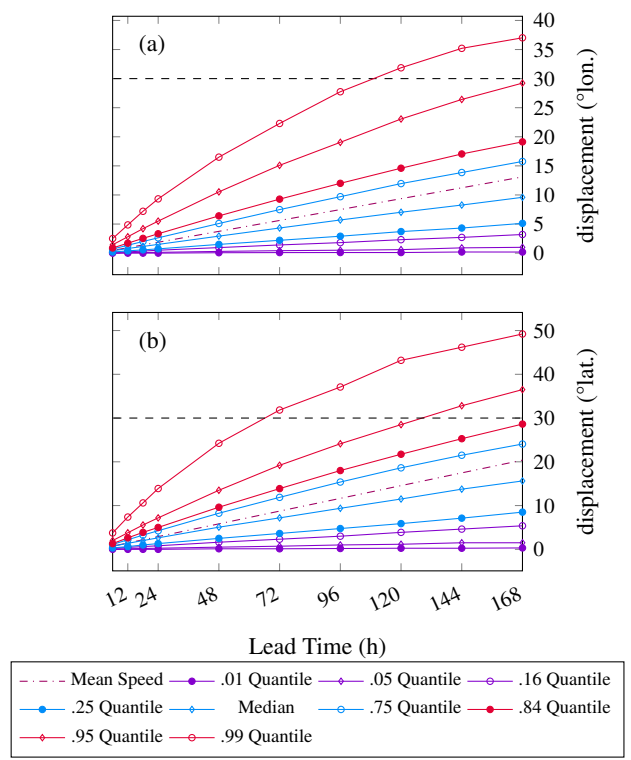
\begin{figure}[h]
    \centering
    \begin{tikzpicture}
        \begin{groupplot}[
            group style={
                group name=Per Lead-Time Curves,
                group size=1 by 2,
                x descriptions at=edge bottom,
                y descriptions at=edge right,
                vertical sep=.5cm,
            },
            footnotesize,
            width=7cm,
            height=5cm,
        ]
        
        \nextgroupplot[
            title={(a)},
            title style={ 
                at={(0.05,0.95)}, 
                anchor=north west, 
                inner sep=2pt 
            },
            legend columns=4,
            legend style={font=\scriptsize, at={(0.5,-0.25)}, anchor=north},
            xmin=6,
            xmax=168,
            ylabel=displacement (\textdegree lon.),
            xtick={12, 24, 48, 72, 96, 120, 144, 168},
            xticklabel style={rotate=25, anchor=north east},
        ]
            \addplot[
                color=black,
                dashed
            ] 
            coordinates {(0.0, 30.0) (168.0, 30.0) };
            \addplot[
                color=jazzberry_jam,
                dashdotted
            ] 
            coordinates {(0.0, 0.0) (8.842105263157896, 0.6896858789307991) (17.68421052631579, 1.3793717578615983) (26.526315789473685, 2.069057636792397) (35.36842105263158, 2.7587435157231965) (44.21052631578948, 3.4484293946539957) (53.05263157894737, 4.138115273584794) (61.89473684210527, 4.827801152515593) (70.73684210526316, 5.517487031446393) (79.57894736842105, 6.207172910377191) (88.42105263157896, 6.896858789307991) (97.26315789473685, 7.58654466823879) (106.10526315789474, 8.276230547169588) (114.94736842105264, 8.965916426100389) (123.78947368421053, 9.655602305031186) (132.63157894736844, 10.345288183961987) (141.47368421052633, 11.034974062892786) (150.31578947368422, 11.724659941823584) (159.1578947368421, 12.414345820754383) (168.0, 13.104031699685182) };
            \addplot[
                color=french_violet,
                mark=*,
                mark size=1.2pt,
            ] 
            coordinates {(6, 0.0) (12, 0.0) (18, 0.0) (24, 0.0) (48, 0.0703125) (72, 0.09375) (96, 0.1015625) (120, 0.10171874999999997) (144, 0.1953125) (168, 0.188359375) };
            
            \addplot[
                color=french_violet,
                mark=diamond,
                mark size=1.2pt,
            ] 
            coordinates {(6, 0.0) (12, 0.09375) (18, 0.1015625) (24, 0.1328125) (48, 0.296875) (72, 0.421875) (96, 0.5) (120, 0.6015625) (144, 0.90625) (168, 1.0) };
            
            \addplot[
                color=french_violet,
                mark=o,
                mark size=1.2pt,
            ] 
            coordinates {(6, 0.1015625) (12, 0.21875) (18, 0.3515625) (24, 0.5) (48, 0.953125) (72, 1.40625) (96, 1.8046875) (120, 2.3125) (144, 2.703125) (168, 3.20125) };
            
            \addplot[
                color=dodger_blue,
                mark=*,
                mark size=1.2pt,
            ] 
            coordinates {(6, 0.1953125) (12, 0.390625) (18, 0.5625) (24, 0.734375) (48, 1.5) (72, 2.203125) (96, 2.90625) (120, 3.703125) (144, 4.3125) (168, 5.12109375)   };
            
            \addplot[
                color=dodger_blue,
                mark=diamond,
                mark size=1.2pt,
            ] 
            coordinates {(6, 0.3984375) (12, 0.796875) (18, 1.1484375) (24, 1.5) (48, 2.921875) (72, 4.3203125) (96, 5.69921875) (120, 7.02734375) (144, 8.265625) (168, 9.59375) };
            
            \addplot[
                color=dodger_blue,
                mark=o,
                mark size=1.2pt,
            ] 
            coordinates {(6, 0.703125) (12, 1.359375) (18, 2.0) (24, 2.6328125) (48, 5.09375) (72, 7.48046875) (96, 9.703125) (120, 11.953125) (144, 13.85546875) (168, 15.7578125) };

            \addplot[
                color=alizarin_crimson,
                mark=*,
                mark size=1.2pt,
            ] 
            coordinates {(6, 0.890625) (12, 1.7134374999999977) (18, 2.5390625) (24, 3.328125) (48, 6.402187499999997) (72, 9.28125) (96, 12.0) (120, 14.599062499999995) (144, 17.046875) (168, 19.128749999999997) };
            
            \addplot[
                color=alizarin_crimson,
                mark=diamond,
                mark size=1.2pt,
            ] 
            coordinates {(6, 1.4375) (12, 2.8125) (18, 4.203125) (24, 5.5) (48, 10.530468749999983) (72, 15.09375) (96, 19.046875) (120, 23.056249999999977) (144, 26.416406249999987) (168, 29.225781249999997) };

            \addplot[
                color=alizarin_crimson,
                mark=o,
                mark size=1.2pt,
            ] 
            coordinates {(6, 2.5023437499999943) (12, 4.852187500000014) (18, 7.193593749999991) (24, 9.327812499999993) (48, 16.5) (72, 22.289375000000007) (96, 27.7421875) (120, 31.859218749999997) (144, 35.204687500000006) (168, 37.015156249999954) };

        \nextgroupplot[
            title={(b)},
            title style={ 
                at={(0.05,0.95)}, 
                anchor=north west, 
                inner sep=2pt 
            },
            legend columns=4,
            legend style={font=\scriptsize, at={(0.5,-0.35)}, anchor=north},
            xmin=6,
            xmax=168,
            ylabel=displacement (\textdegree lat.),
            xtick={12, 24, 48, 72, 96, 120, 144, 168},
            xticklabel style={rotate=25, anchor=north east},
            xlabel=Lead Time (h)
        ]
            \addplot[
                color=jazzberry_jam,
                dashdotted
            ] 
            coordinates {(0.0, 0.0) (8.842105263157896, 1.072764645435563) (17.68421052631579, 2.145529290871126) (26.526315789473685, 3.2182939363066887) (35.36842105263158, 4.291058581742252) (44.21052631578948, 5.363823227177815) (53.05263157894737, 6.436587872613377) (61.89473684210527, 7.50935251804894) (70.73684210526316, 8.582117163484504) (79.57894736842105, 9.654881808920067) (88.42105263157896, 10.72764645435563) (97.26315789473685, 11.800411099791193) (106.10526315789474, 12.873175745226755) (114.94736842105264, 13.945940390662319) (123.78947368421053, 15.01870503609788) (132.63157894736844, 16.091469681533447) (141.47368421052633, 17.16423432696901) (150.31578947368422, 18.23699897240457) (159.1578947368421, 19.309763617840133) (168.0, 20.382528263275695) };
            \addlegendentry{Mean Speed}
            
            \addplot[
                color=french_violet,
                mark=*,
                mark size=1.2pt,
            ] 
            coordinates {(6, 0.0) (12, 0.0) (18, 0.0) (24, 0.0) (48, 0.125) (72, 0.125) (96, 0.1875) (120, 0.25) (144, 0.25) (168, 0.31937499999999996) };
            \addlegendentry{.01 Quantile}
            
            \addplot[
                color=french_violet,
                mark=diamond,
                mark size=1.2pt,
            ] 
            coordinates {(6, 0.0625) (12, 0.125) (18, 0.1875) (24, 0.25) (48, 0.5) (72, 0.7140625000000007) (96, 1.0) (120, 1.1593750000000007) (144, 1.5) (168, 1.5)  };
            \addlegendentry{.05 Quantile}
            
            \addplot[
                color=french_violet,
                mark=o,
                mark size=1.2pt,
            ] 
            coordinates {(6, 0.21875) (12, 0.4375) (18, 0.625) (24, 0.8125) (48, 1.625) (72, 2.3125) (96, 3.0) (120, 3.875) (144, 4.625) (168, 5.375) };
            \addlegendentry{.16 Quantile}
            
            \addplot[
                color=dodger_blue,
                mark=*,
                mark size=1.2pt,
            ] 
            coordinates {(6, 0.3125) (12, 0.625) (18, 1.0) (24, 1.3125) (48, 2.5) (72, 3.625) (96, 4.75) (120, 5.875) (144, 7.125) (168, 8.5) };
            \addlegendentry{.25 Quantile}
            
            \addplot[
                color=dodger_blue,
                mark=diamond,
                mark size=1.2pt,
            ] 
            coordinates {(6, 0.6875) (12, 1.375) (18, 2.0) (24, 2.625) (48, 5.0625) (72, 7.1875) (96, 9.375) (120, 11.5) (144, 13.75) (168, 15.625) };
            \addlegendentry{Median}
            
            \addplot[
                color=dodger_blue,
                mark=o,
                mark size=1.2pt,
            ] 
            coordinates {(6, 1.125) (12, 2.1875) (18, 3.25) (24, 4.25) (48, 8.25) (72, 11.875) (96, 15.375) (120, 18.625) (144, 21.5) (168, 24.0625) };
            \addlegendentry{.75 Quantile}

            \addplot[
                color=alizarin_crimson,
                mark=*,
                mark size=1.2pt,
            ] 
            coordinates {(6, 1.3125) (12, 2.625) (18, 3.875) (24, 5.0) (48, 9.625) (72, 13.875) (96, 18.0) (120, 21.72999999999996) (144, 25.28125) (168, 28.625)  };
            \addlegendentry{.84 Quantile}
            
            \addplot[
                color=alizarin_crimson,
                mark=diamond,
                mark size=1.2pt,
            ] 
            coordinates {(6, 2.0) (12, 3.78125) (18, 5.560937499999966) (24, 7.1875) (48, 13.5) (72, 19.206249999999955) (96, 24.125) (120, 28.5) (144, 32.8125) (168, 36.5) };
            \addlegendentry{.95 Quantile}

            \addplot[
                color=alizarin_crimson,
                mark=o,
                mark size=1.2pt,
            ] 
            coordinates {(6, 3.75) (12, 7.375) (18, 10.605937499999982) (24, 13.890312499999993) (48, 24.23531250000002) (72, 31.83687500000002) (96, 37.09375) (120, 43.19687499999952) (144, 46.18828125000003) (168, 49.21132812499995)};
            \addlegendentry{.99 Quantile}

            \addplot[
                color=black,
                dashed
            ] 
            coordinates {(0.0, 30.0) (168.0, 30.0) };
            
        \end{groupplot}
    \end{tikzpicture}
    
    \caption{Training Set Quantiles for \mbox{(a) Zonal} Displacement and \mbox{(b) Meridional} Displacement} 
    \label{SIfig:leadtime_displacements}
\end{figure}

Here we seek to define a spatial domain for the input features that will be used by our post-processing algorithms, centered at the position of the storm at the time of forecast (i.e., $t_0$). To do with, we first calculate storm displacements across all lead times of interest for all the storms in the training set. We then estimate the following quantile levels: 0.01, 0.05, 0.16, 0.25, 0.5, 0.75, 0.84, 0.95, and 0.99 for zonal and meridional storm displacements. As shown in \Cref{SIfig:leadtime_displacements}, a fixed $\pm30^\circ$ box centered on the storm's initial location captures the center of over 80\% of storms at the 7-day lead time and over 95\% at 5 days lead time, which is the typical operational forecast horizon for TCs, and we thus define the spatial domain as the $\pm30^\circ$ box centered at the position of the TC at the time of forecast (i.e., $t_0$). Importantly, we choose a fixed size domain in order to ensure that we can process the fields with common deep learning algorithms (e.g., Convolutional Neural Networks--CNNs--- with dense layers), though methods for training algorithms that generalize across domain sizes exist (\mbox{e.g., fully} convolutional networks and neural operators). While our definition of the domain size can be considered a hyperparameter and optimized for best performance, we deem the size adequate for purposes and do not optimize it further.

\tikzsetnextfilename{Figure_2}

\subsubsection{Masking} Large and small scale features may be important to predicting the intensity of storms. For example, it has been shown that teleconnections play an important role in subseasonal and season predictions and modeling of tropical cyclones~\citep{Domeisen22Advances, patricola2017teleconnection, bell2014simulation, feng2020western}. However, we theorized that our choice of domain size (i.e., $\pm30^\circ$) could pose a challenging problem--e.g., given cases where more than one TC is present in the domain for a number of timestamps. As an example of this, we can point to hurricane Maria (2017), that coexisted in the Caribbean alongside hurricanes Jose and Lee.

\begin{figure}[h]
    \centering
    \begin{subfigure}[]{0.3\textwidth}
        \centering
        \includegraphics[width=.95\textwidth]{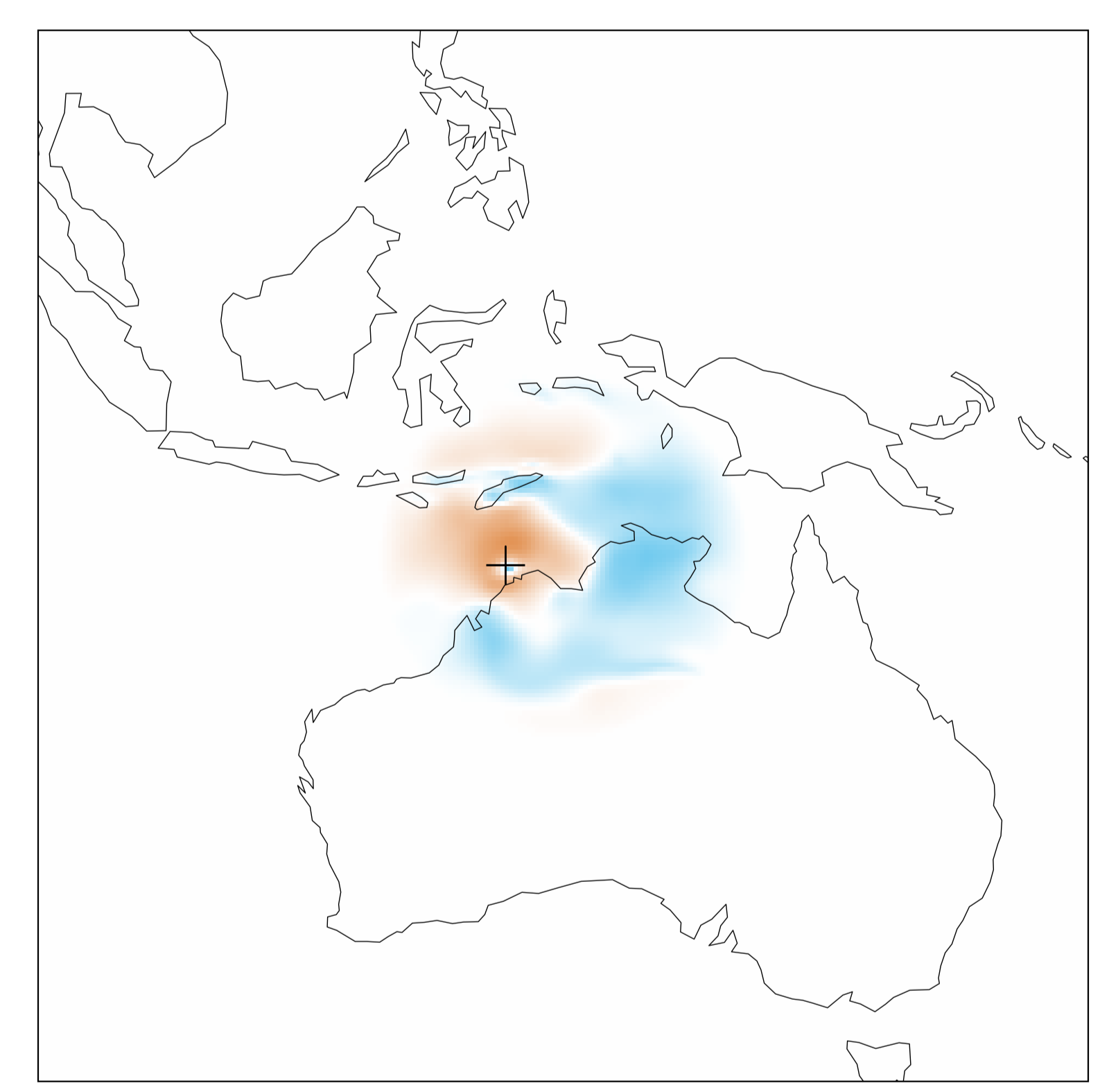}
        \caption{Lead time 18h}
        \label{fig:masked_18}
    \end{subfigure}
    \hfill\\
    \begin{subfigure}[]{0.3\textwidth}
        \centering
        \includegraphics[width=.95\textwidth]{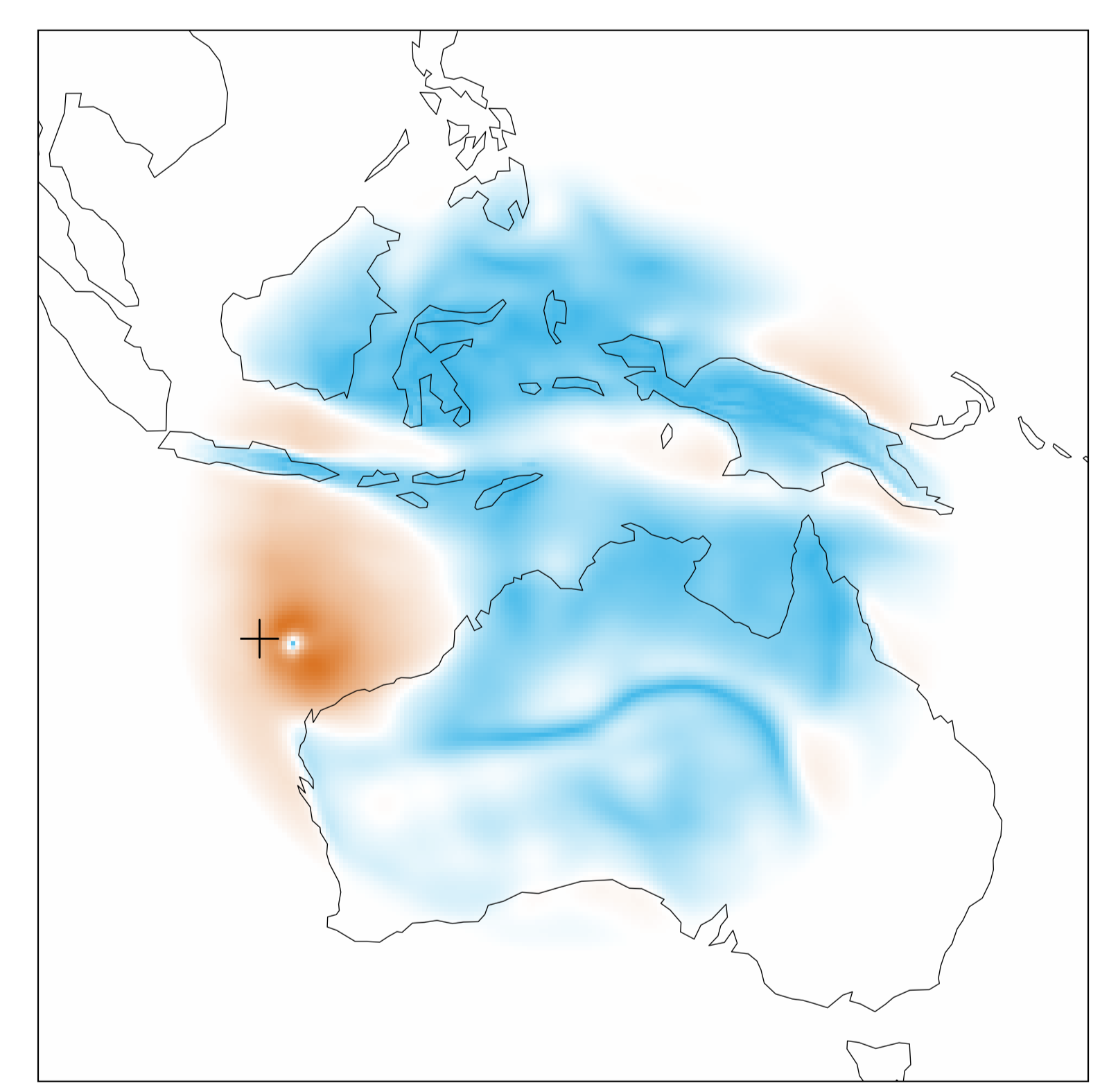}
        \caption{Lead time 72h}
        \label{fig:masked_72}
    \end{subfigure}\\
    \begin{subfigure}[]{0.3\textwidth}
        \centering
        \includegraphics[width=.95\textwidth]{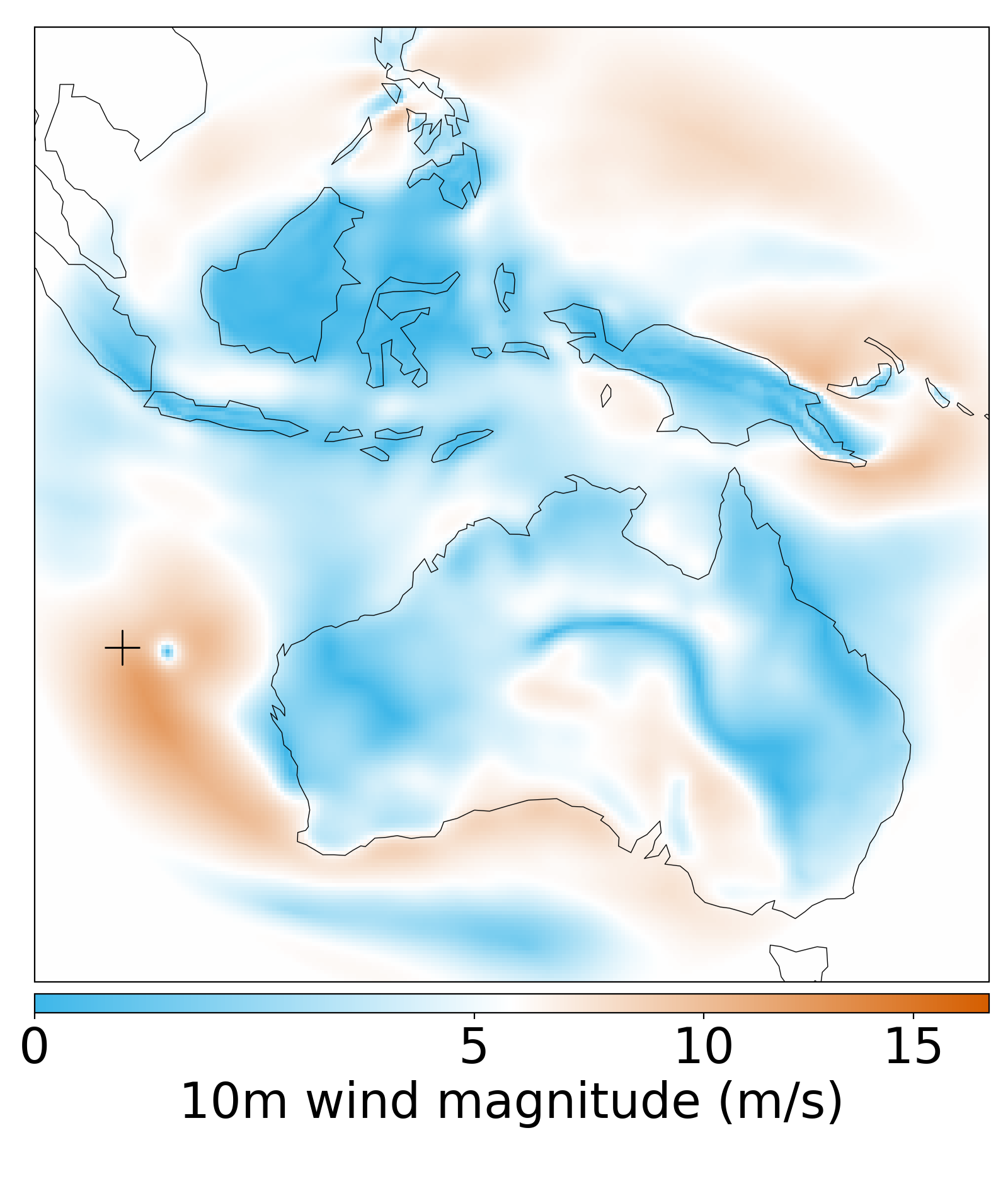}
        \caption{Lead time 120h}
        \label{fig:masked_120}
    \end{subfigure}
    \caption{Masked wind magnitude field for the PanguWeather forecast at initial time 2020-01-11 00h00 (associated with TC Claudia) for different lead times. The black plus ($+$) symbols indicate the position of TC Claudia at the given lead time according to IBTrACS.}
    \label{fig:masking}
\end{figure}

To assuage the impacts of the large domain, we propose the use of a mask to effectively reduce the domain being considered by the post-processing algorithm at shorter lead times while retaining a consistent spatial domain as is required by the neural network architectures we use in our study. Thus, we investigate the use of a similar approach to that used when defining the domain size and propose the use of a mask with a radius corresponding to the 0.84 quantile of displacement (in km) for the storms in our training dataset. The masking radius is further increased by a linear fade out to 300km beyond the displacement, which corresponds to approximately 2 times the 0.99 quantile of the radius of maximum wind of observed storms as reported by~\cite{chavas2022simple}. The area of the fields outside the mask is then set to the mean value for that field in the training dataset. A sample of the resulting fields is given in \Cref{fig:masking}, where it can be seen that though the overall domain is of a constant size the area exhibiting variability in field values becomes progressively larger with an increase in lead time. We test the use of both masked and unmasked NeWM fields as inputs to our prediction algorithms, and provide a sensitivity analysis in Section S-1--further noting that the mask's parameters are hyperparameters that can be optimized.
While the grid spacing is constant in degrees, we know that the km value varies depending on the distance from the equator. However, given that we are working in the tropics we consider the grid spacing to be relatively constant and to be $1^\circ \approx 100\text{km}$ -- allowing us to use masks that are conditioned only on the lead time.

\subsection{Normalization} After preparing the mask and unmasked fields, we proceed to scale them using the Standard Scaling (i.e., z-score scaling) method, which subtracts the mean value ($\mu$) and divides by the standard deviation ($\sigma$) for each field. We note that this scaling centers the distribution of each field at 0 and makes 1 correspond to one standard deviation. This method shows some sensitivity to outliers and is well-established for normally-distributed data. We find it to be a sufficiently robust scaling for our purposes as it transforms the values of each feature space to comparable, unitless values while preserving the underlying shape of each feature distribution. Consider the model output vector \( \mathbf{o}_i \) representing the $i^{\text{th}}$ model output sample in our dataset:
\begin{equation}
    \mathbf{o}_i = \begin{pmatrix}
    \left| \mathbf{V}_{10m, \,i} \right|\\
    \mathbf{\vartheta}_{10m, \,i}\\
    \mathbf{P}_{0, \,i} \\
    \mathbf{z}_{500,\, i} \\
    \mathbf{T}_{850,\, i} \\
    \end{pmatrix}
\end{equation}
The scaled fields are given by 
\begin{equation}
    \mathbf{o}_{i,scaled} =  \frac{\mathbf{o}_i - \mathbf{{\mu}_{train}}}{\mathbf{{\sigma}_{train}}},
\end{equation}
where \( \boldsymbol{\mu}_{\text{train}} \) and \( \boldsymbol{\sigma}_{\text{train}} \) are the mean and standard deviation vectors computed over the entire training dataset. For consistency, these statistics are calculated separately for each variable and used to normalize all samples in the training, validation, and test sets. 

\subsection{Data Split}
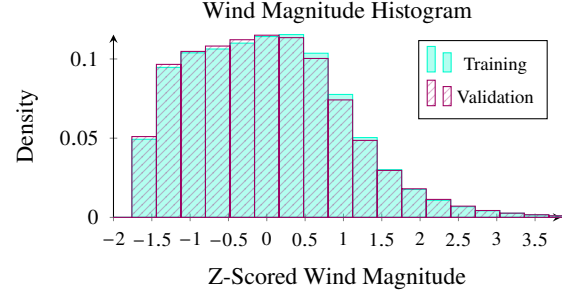
\begin{figure}[h]
\centering
    \begin{tikzpicture}
        \begin{axis}[
            ybar,
            footnotesize,
            width=7.5cm,
            height=4cm,
            xticklabel style={font=\scriptsize},
            xlabel={Z-Scored Wind Magnitude},
            yticklabel style={
                /pgf/number format/fixed,
                /pgf/number format/precision=2
            },
            axis lines=left,
            title={Wind Magnitude Histogram},
            ylabel={Density},
            legend pos=north east,
            legend style={font=\scriptsize},
            ytick={0,.05,.10,.15},
            xmin=-2,
        ]
        \addplot[
            color=aquamarine,
            fill,
            fill opacity=0.3,
            bar width=0.32,
            bar shift=0pt,
        ] 
        table [x=Train_Bin_Centers, y=Train_Density, col sep=comma]{csvs/H1-Magnitude_histogram_unmasked.csv};
        \addlegendentry{Training}
        \addplot[
            color=jazzberry_jam,
            fill,
            fill opacity=0.3,
            bar width=0.32,
            bar shift=0pt,
            pattern=north east lines,
            pattern color=jazzberry_jam,
        ] 
        table [x=Valid_Bin_Centers, y=Valid_Density, col sep=comma]{csvs/H1-Magnitude_histogram_unmasked.csv};
        \addlegendentry{Validation}
        \end{axis}
    \end{tikzpicture}
    \caption{Wind Magnitude feature pixel-wide distribution on the training and validation set (unmasked inputs). For all feature distributions, refer to Figure S-2. 
    }
    \label{fig:distSample}
\end{figure}

For tractability, we fix the years used to train the models across all experiments: 2013-2017 were used in the training, 2018 and 2019 were used for validation, and 2020 was left as an unseen test set. We highlight that by year we refer to the calendar year and not the cyclone season, as some seasons span multiple calendar years (e.g., the tropical cyclone season in the Australian basin). We further visually verified that the pixel-wise distribution of the calculated fields overlap significantly when comparing the feature distributions in the training set and the validation set, as shown in \Cref{fig:distSample}.

\tikzsetnextfilename{Figure_4}

\subsection{Summary}

After each forecast is performed we retrieve the starting position of the TC from IBTrACS. We then find the closest grid point to that location using the Haversine distance~\citep{inman1849navigation}, and clip the region determined by a square centered at the grid point and of size 60 degrees in latitude and longitude. This results in NeWM fields with dimensions $241 \times 241 \times c$, where $c$ is the number of variables (5 for this study). We additionally produce fields that have been masked so that the values outside a radius are set to the mean value of each field in the training dataset. This radius increases with lead time. The fields are then scaled using standard scaling with the mean and standard deviation calculated for each field using training data.

\section{Post-Processing Algorithms}

To the best of our knowledge, the problem of post-processing the outputs of AI weather models for TC intensity prediction has not yet been addressed. We thus explore a number of approaches to post-process such outputs, ranging from linear to deep learning models.

\subsection{Inputs}
\paragraph{Linear Models:} To train a simple and interpretable multiple linear regression (MLR) baseline, we summarize NeWM-predicted fields using spatial statistics. In addition to these summary statistics, we include the forecast lead time ($\tau_i$), the observed maximum wind speed ($V_{\mathrm{max},\,i}$), and the observed minimum sea-level pressure ($P_{\mathrm{min},\,i}$) at the time of forecast. The input vector for the $i^{\text{th}}$ sample, denoted by \( \mathbf{x}_{\text{lin},\,i} \), is defined as:
\begin{equation}
    \mathbf{x}_{\text{lin},\,i} = \begin{pmatrix}
        \max\left( \left| \mathbf{V}_{10m, \,i} \right| \right) \\
        \min\left( \mathbf{P}_{0, \,i} \right) \\
        \mathrm{range}\left( \left| \mathbf{V}_{10m, \,i} \right| \right) \\
        \mathrm{range}\left( \mathbf{P}_{0, \,i} \right) \\
        \min\left( \mathbf{z}_{500,i} \right) \\
        \mathrm{range}\left(\mathbf{T}_{850,i} \right) \\
        \tau_i \\
        V_{max,\,i} \\
        P_{min,\,i}
    \end{pmatrix},
    \label{eq:linInputs}
\end{equation}
where the operations min, max, and range are applied spatially over each field.

\paragraph{Artificial Neural Networks (ANNs):} In this study, we use feedforward dense ANNs to capture non-linear relationships in the data used to train linear models--thereby adding a small degree of complexity. The inputs are thus the same as for the linear models (i.e., $\mathbf{x}_{lin}$).  
\paragraph{Convolutional Neural Networks (CNNs):} We use CNNs to test how well algorithms are able to learn from the spatial information present in NeWM outputs. Thus, instead of summarizing the fields as was done for $\mathbf{x}_{lin}$, we use the fields themselves. In addition to the fields, we prepare a vector of scalars that provide important context to the algorithm. The vector includes the forecast's lead time ($\tau_i$), the maximum wind observed at the time of forecast ($V_{max,\,i}$), the minimum sea level pressure observed at the time of forecast ($P_{min,\,i}$), and the central latitude and longitude observed at the time of forecast ($\theta$ and $\phi$, respectively). We therefore define the field inputs $\mathbf{x}_{\mathrm{CNN, \,i}}$ and the vector inputs $\mathbf{x}_{vec, \,i}$ for the $i^{\text{th}}$ sample as follows:
\begin{equation}
    \mathbf{x}_{\mathrm{CNN, \,i}} = \mathbf{o}_i = \begin{pmatrix}
    \left| \mathbf{V}_{10m, \,i} \right|\\
    \mathbf{\vartheta}_{10m, \,i}\\
    \mathbf{P}_{0, \,i} \\
    \mathbf{z}_{500,\, i} \\
    \mathbf{T}_{850,\, i} \\
    \end{pmatrix}, \qquad \mathbf{x}_{vec, \,i} =\begin{pmatrix}
            \tau_i\\
            V_{max,\,i}\\
            P_{min,\,i}\\
            \theta_i\\
            \phi_i
            \end{pmatrix}.
\end{equation}
\subsection{Outputs}
As mentioned in subsection~\ref{sec:Dataset}\ref{subsec:ibtracs}, we rely on IBTrACS to generate the ground truth that we target with our algorithms. To facilitate learning, instead of attempting to predict the reported value of intensity--maximum wind $V_{max}$ and minimum pressure $P_{min}$--- we attempt to predict the \textit{intensification}: $\Delta V_{max}=V_{max}\left(t+\tau\right) - V_{max}\left(t\right)$ and $\Delta P_{min}= P_{min}\left(t+\tau\right) - P_{min}\left(t\right)$. This is a choice that we make because we expect the distribution of intensification values to be fairly Gaussian and thus simpler to model probabilistically using distributional regression methods, as will be described in following subsections. 
\subsection{Loss Functions}
All algorithms in our study are trained to minimize a loss function, including the linear models. While the deterministic linear models could be fit using the normal equation, we decided to fit it with an optimizer and deterministic loss both to (1) maintain a consistent data pipeline across deterministic and probabilistic linear models, and (2) ensure that the fitting of linear and non-linear models is consistent. We thus rely on two loss functions for fitting our models: (1)) the Mean Squared Error (MSE) for deterministic predictions (i.e., when we produce a single intensity forecast for a given set of initial conditions and lead time) and (2) the Continuous Ranked Probability Score (CRPS) for probabilistic predictions (i.e., when we predict a distribution of possible intensities for a given set of initial conditions and lead time).
\paragraph{MSE:} The MSE is a statistical measure used to evaluate the model performance by calculating the average of the squares of the errors between the observed values ($\mathbf{y}$) and predicted values ($\mathbf{{\hat y}}$): 
\begin{equation}
    \text{MSE}\left(\mathbf{y}, \mathbf{\hat{y}} \right) = \frac{1}{n}\sum_{i=0}^{n}\left( \mathbf{y}_i-{\mathbf{{\hat y}}_i} \right)^2.
\end{equation}
\paragraph{CRPS:} The CRPS is a proper score used to evaluate the distance between two distributions, and can be defined as the area between the cumulative distribution of the distributions being compared. 
Given that we will be targeting intensification, we can instead rely on the closed form solution of the CRPS between our prediction of a gaussian distribution with predicted mean $\hat{\mu}$ and variance $\hat{\sigma}^2$---i.e., $\mathcal{N}\left( \hat{\mu}, \, \hat{\sigma}^2\right)$--- and the degenerate distribution of the observation ~\citep{gneiting2005calibrated}:
\begin{equation}
    \text{CRPS}\left[y, \hat{\mu}, \hat{\sigma} \right] = \hat{\sigma}\left\{ \tilde{e} \left[2 \Phi \left( \tilde{e} \right) - 1 \right] + 2\varphi \left( \tilde{e} \right) - \frac{1}{\sqrt{\pi}}\right\}\\    
\end{equation}\\
where $\tilde{e} =(y-\hat{\mu})/\hat{\sigma}$ is the standardized prediction error, $\varphi \left( \tilde{e} \right)$ is the PDF of the normal Gaussian distribution with mean 0 and variance 1 evaluated at $\tilde{e}$, and $\Phi\left( \tilde{e} \right)$ is the CDF of that same distribution evaluated at $\tilde{e}$.

\subsection{Linear Models}
For linear models, we rely on multiple linear regression algorithms, which are a simple linear combination of the inputs plus a bias:
\begin{equation}
    \hat{y} = \alpha_0 + \alpha_1 x_1 + \alpha_2 x_2 + \cdots + \alpha_n x_n,
    \label{eq:linModel}
\end{equation}
where $\alpha_0$ is a learned bias term and $\{\alpha_1, \alpha_2, ... \alpha_n\}$ are the learned coefficients of $n$ features in input vector $\mathbf{x}$.
\par For the deterministic setup, we train two independent MLR algorithms: one that targets $\Delta V_{max}$ and one that targets $\Delta P_{min}$. 
For the probabilistic setup, we instead train two pairs of MLR algorithms. The first pair of MLRs is used to target the mean and standard deviation of wind intensification \mbox{(i.e., $\mu_{\Delta V_{max}}$} and $\sigma_{\Delta V_{max}}$, respectively) while the second pair targets the mean and standard deviation of pressure intensification \mbox{(i.e., $\mu_{\Delta P_{min}}$} and $\sigma_{\Delta P_{min}}$). We emphasize that for our distributional regression setup we assume that the predicted distribution is Gaussian, and thus evaluate our prediction with the closed form solution of the CRPS between a Gaussian distribution and an observation.
\subsection{Artificial Neural Networks}
In our study, we use a simple multilayer perceptron architecture across all experiments that includes 6 layers containing a number of neurons set to 4 times the number of features in a sample. We additionally set the activation function to be consistent across all layers, and train models using the HardSwish function and leaky ReLU function.
\par For the deterministic setup, a final layer comprising two units is used after the aforementioned 6 layers. The activation function is set to the identity function, and the network targets $\Delta V_{max}$ and $\Delta P_{min}$. For the probabilistic setup, we instead target two normal distributions and thus use a final layer comprising 4 neurons with the identity activation function (used to predict $\mu_{\Delta V_{max}}, \sigma_{\Delta V_{max}}, \mu_{\Delta P_{min}}\text{, \;and} \ \ \sigma_{\Delta P_{min}}$). 

\subsection{Convolutional Neural Networks}

While it is possible to set up convolutional architectures that are able to process fields of varying spatial resolutions, we opt to use dense layers (similar to those described for ANNs, above) after the convolutional layers to target the intensification of the storm. Thus, in the deterministic setup the final layer continues to be a layer with two neurons that target  $\Delta V_{max}$ and $\Delta P_{min}$. Similarly, the probabilistic setup performs a distributional regression that targets two normal distributions and the final layer thus comprises 4 neurons that predict $\mu_{\Delta V_{max}}, \sigma_{\Delta V_{max}}, \mu_{\Delta P_{min}}\text{, \;and} \ \ \sigma_{\Delta P_{min}}$. An overview of the architectures and hyperparameter searches conducted for the CNNs is given below.

\paragraph{``Vanilla'' CNNs:}
In this study, we set up vanilla CNNs comprising 3 convolutional layers and 4 dense layers. The spatial field inputs $\mathbf{x}_{\mathrm{CNN}}$ are processed by the three convolutional layers sequentially (with max pooling layers applied after each convolutional layer), after which the feature maps of the third convolutional layer are flattened and used as inputs to the first dense layer. The vector inputs $\mathbf{x}_{vec}$ are encoded by the second and third dense layers. Then the outputs of the first dense layer (the branch that processed $\mathbf{x}_{\mathrm{CNN}}$) and third dense layer are concatenated and processed by the fourth and final dense layer.
\par We ran experiments with additional layers, including batch normalization layers in between the convolutional and pooling layers, dropout2d layers after the pooling layers, and dropout layers after each dense layer. A diagram illustrating the architectures can be found in Figure S-3.

\paragraph{UNet:}
In addition to the vanilla CNNs described above, we train simple UNets as a way of making the architecture more expressive and trying to improve generalization. UNets use a combination of reduction to a latent space with convolutional layers (A brief introduction to UNets is provided by ~\cite{chase2023machine}, and we point the reader to the paper where UNets were introduced ~\citep{ronneberger2015u} should more details be required. We note that one of the main advantages of using UNets stems from its use of skip connections between the encoder and decoder layers - which allow the network to better preserve spatial information that may be lost in latent representations. 
\par In this study, we use UNets comprising three encoding convolutional layers, one bottleneck layer, three decoding convolutional transpose layers, and four dense layers. After the decoding layers, the flattened feature map is used as input to a fully connected layer. The vector inputs $\mathbf{x}_{vec}$ are handled in the same way as in the vanilla CNNs, and the output of the first dense layer is again concatenated with the output of the third dense layer before processing by the fourth and final dense layer. A diagram illustrating the architectures can be found in Figure S-4.

\subsection{Hyperparameter Tuning:}
Here we present an overview of the hyperparameter space searched, summarized in \Cref{tab:hp_tuning}. We note that the spaces searched were empirically selected.

\begin{table}[h]
    \centering
    \begin{tabular}{ccc}
          \shortstack{Model\\Architecture}& Hyperparameter& Search Space\\
          \hline
        \multirow{14}{*}{\centering\shortstack{Model\\Agnostic}}& \multirow{2}{*}{Wind Vars.}&  $\{ \left\{ u_{10m}, v_{10m} \right\}$,\\
             & &$\left\{ \left| V_{10m} \right|, \vartheta_{10m} \right\}\}$\\
             & \multirow{3}{*}{Feature Ablation}& $\{\left\{Z_{500}\right\},$ \\
             & & $\left\{T_{850}\right\},$\\
             & & $ \left\{ Z_{500}, T_{850} \right\}\}$\\
             & \multirow{4}{*}{\centering$\tau$} & \{1 Model per $\tau$, \\
             & & $6h \leq \tau \leq 24h$, \\
             & & $48h \leq \tau \leq 168h$, \\
             & & all $\tau \}$\\
             & \multirow{2}{*}{\centering target} & $\{ \left\{ V_{max},\; P_{min} \right\},$\\
             & & $\left\{ \Delta V_{max},\; \Delta P_{min} \right\} \}$ \\
             & \multirow{3}{*}{\centering \shortstack{Learning Rate\\ Schedule}}& \{Exponential,\\
             && Cyclical,\\
             && None\}\\
             &&\\
             &&\\
        ANN & Activation Function &  \{ Swish, Relu, Tanh\}\\
             &&\\
             &&\\
        \multirow{6}{*}{\centering\shortstack{CNN}}& Batch Normalization&  \{with, without\}\\
             & \multirow{3}{*}{\centering Layer Widths} & $\{ \left[ 8,16,32 \right],$\\
             && $\left[ 32,64,128 \right],$\\
             && $\left[ 64,128,256 \right] \}$\\
             & Dropout Rate ($\delta$) & \{ $ 0.025 \leq \delta \leq 0.975 $  \}\\
             & L2 Regularization & \{ $0,\; .001 $  \}\\
             &&\\
             &&\\
        \multirow{4}{*}{\centering\shortstack{UNet}}& Batch Normalization&  \{with, without\}\\
             & \# of UNet Channels & $\left\{ 1,\;4,\;32 \right\},$\\
             & Dropout Rate & $ \left\{ 0.33,\; 0.5,\; 0.66\right\} $\\
             & L2 Regularization & \{ $0,\; .001$  \}\\
             &&\\
             &&\\
    \end{tabular}
    \caption{Hyperparameter Search Space}
    \label{tab:hp_tuning}
\end{table}

\subsection{Baseline Models}
We further provide simple baselines for comparison against the performance of the studied algorithms. An overview of each baseline is given below.
\paragraph{Persistence:}
The first baseline we evaluate against is persistence, which given our problem statement (i.e., predicting the intensification of a known tropical system) requires that we predict 0 intensity changes. This is a fairly good guess at short lead times, but the quality of the prediction decays quickly with an increase in lead time.
\paragraph{Average Climatology:}
The second baseline we propose is to calculate the average behavior of storms at a given lead time. This results in the climatology values shown in \Cref{tab:ave_climatology}, shown in z-score values. We note that the distributions are generally centered around 0, and that as expected the variance increases with lead time---with the notable exception of 168h. A rigorous examination of this variance is not within the scope of this study, but we propose that part of the decrease in variance at maximum lead time is related to our choice to eliminate lead time predictions for which we do not have a ground truth value.
\begin{table}[h]
    \centering
    \begin{tabular}{rrlrl}
         Lead Time& $\mu_{\Delta V_{max}}$& $\sigma_{\Delta V_{max}}$ & $\mu_{\Delta P_{min}}$ & $\sigma_{\Delta P_{min}}$\\
         \hline
           6h &-0.07007&  0.2118& 0.08990 &0.2142 \\
          12h &-0.06155&  0.3794& 0.08030 &0.3818 \\
          18h &-0.04970&  0.5044& 0.06690 &0.5190 \\
          24h &-0.03455&  0.6304& 0.04970 &0.6530 \\
          48h & 0.03640&  1.024 &-0.03330 &1.028  \\
          72h & 0.08856&  1.295 &-0.10364 &1.299  \\
          96h & 0.10700&  1.408 &-0.13700 &1.427  \\
         120h & 0.10077&  1.492 &-0.14280 &1.503  \\
         144h & 0.07996&  1.535 &-0.13270 &1.501  \\
         168h & 0.04697&  1.477 &-0.11273 &1.460  \\

    \end{tabular}
    \caption{Average Climatology per Lead Time}
    \label{tab:ave_climatology}
\end{table}
\paragraph{Direct Regional Forecast (No post-processing):}
Unlike the other baselines, this baseline relies on the \textit{inputs} of the NeWMs used in this study. Rather than providing inputs forecasted by NeWMs to the post-processing architectures, we instead provide the ERA5 conditions at the time of forecast irrespective of lead time (i.e., the data that the NeWMs process in order to provide a forecast). Comparing against this baseline (a direct regional forecast without any intermediate prediction step) helps evaluate the benefit of post-processing NeWMs.

\section{Results}

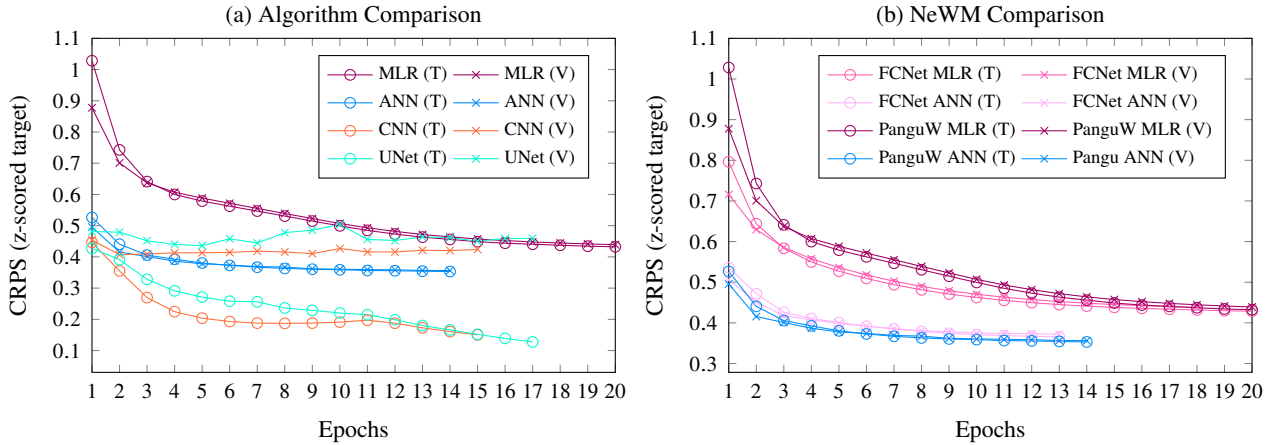
\begin{figure*}[h]
    \centering
    \begin{tikzpicture}
        \begin{groupplot}[
            group style={
                group name=Training Curves,
                group size=2 by 1,
                x descriptions at=edge bottom,
                horizontal sep=1.5cm,
            },
            footnotesize,
            width=8.5cm,
            height=6cm,
            xlabel=Epochs,
            ylabel=CRPS (z-scored target),
        ]
        \nextgroupplot[
            title={(a) Algorithm Comparison},
            legend style={at={(0.95,0.95)}, anchor=north east, legend columns=2, font=\scriptsize},
            legend cell align=left,
            xmin=1,
            xmax=20,
            ymax=1.1
        ]
            \addplot[
                color=jazzberry_jam,
                mark=o
            ] 
            coordinates {(1,1.0282852565064842) (2,0.742680866961126) (3,0.6411708613604675) (4,0.5999477695167801) (5,0.5788145943740268) (6,0.5624155461972142) (7,0.5467426890024433) (8,0.5307433508612491) (9,0.514748517193912) (10,0.49947595108806353) (11,0.4849628197190202) (12,0.47308646032103785) (13,0.4632909075345522) (14,0.4559279416262368) (15,0.4488282389478919) (16,0.44386894522625725) (17,0.44060678798475383) (18,0.43722661924950873) (19,0.4342654308787099) (20,0.4323862984224602) };
            \addlegendentry{MLR (T)}
            \addplot[
                color=jazzberry_jam,
                mark=x
            ] 
            coordinates {(1,0.8777127081012152) (2,0.7007104618003569) (3,0.6370390280183539) (4,0.6071800866040838) (5,0.5882741662572666) (6,0.571657545505518) (7,0.5552659514977272) (8,0.5389635302575237) (9,0.5227860583957419) (10,0.5070663221090673) (11,0.49322115567075203) (12,0.48167703659778616) (13,0.47191499056945363) (14,0.46400632044996126) (15,0.45736368458314114) (16,0.45218662118696307) (17,0.4479710687714887) (18,0.44449475847453956) (19,0.44149557626750097) (20,0.4390308677432049)};
            \addlegendentry{MLR (V)}
            %
            \addplot[
                color=dodger_blue,
                mark=o
            ] 
            coordinates {(1,0.5265016231087991) (2,0.44059723762818326) (3,0.4056626508633296) (4,0.3927132620671649) (5,0.380366372289481) (6,0.3727545980134128) (7,0.36698373039195564) (8,0.3629398860129309) (9,0.360499593394774) (10,0.35877886828449035) (11,0.3565240016690007) (12,0.35566972610023284) (13,0.3542612731272792) (14,0.3530058160792162)  };
            \addlegendentry{ANN (T)}
            \addplot[
                color=dodger_blue,
                mark=x
            ] 
            coordinates {(1,0.4955298553209707) (2,0.41544645333505537) (3,0.40094357486589843) (4,0.38677140854927433) (5,0.37775698250316714) (6,0.3739656871521329) (7,0.36921661679285117) (8,0.36743274052818137) (9,0.3619588200586388) (10,0.3605978534702795) (11,0.3594074409948774) (12,0.3585718707686447) (13,0.3566134505602251) (14,0.35634245032287504)};
            \addlegendentry{ANN (V)}
            %
            \addplot[
                color=outrageous_orange,
                mark=o
            ] 
            coordinates {(1,0.4463214549554847) (2,0.35521586961049867) (3,0.26950683308985063) (4,0.22455077855280337) (5,0.203773660075849) (6,0.19304841851934473) (7,0.188004185977019) (8,0.18710454819107886) (9,0.18801957690220045) (10,0.19104924277768146) (11,0.1971796037150075) (12,0.18764936439188581) (13,0.17299639736833608) (14,0.1608895919799113) (15,0.15029157889534933)};
            \addlegendentry{CNN (T)}
            \addplot[
                color=outrageous_orange,
                mark=x
            ] 
            coordinates {(1,0.4538623960414171) (2,0.4067992384708117) (3,0.4095818676105162) (4,0.41307617128683777) (5,0.413266670091172) (6,0.4141791880310978) (7,0.41855831073142324) (8,0.41571996921359455) (9,0.4105120489397737) (10,0.4268140407759226) (11,0.4160612007196168) (12,0.4159089563779099) (13,0.4213191606526742) (14,0.42025824446436677) (15,0.42378246813494425)};
            \addlegendentry{CNN (V)}
            %
            \addplot[
                color=aquamarine,
                mark=o
            ] 
            coordinates {(1,0.42727046723247775) (2,0.39122199624984283) (3,0.3281963661785561) (4,0.2911726522957366) (5,0.27119313738231776) (6,0.25814006247680865) (7,0.25610194892866706) (8,0.23636210716148726) (9,0.22883590653665473) (10,0.2204122666531723) (11,0.21469555421799352) (12,0.1986918550586848) (13,0.18002168179618602) (14,0.1663411598528713) (15,0.15194477385798813) (16,0.13895897729767448) (17,0.1274817848933914) };
            \addlegendentry{UNet (T)}
            \addplot[
                color=aquamarine,
                mark=x
            ] 
            coordinates {(1,0.481098059756158) (2,0.4792952423864621) (3,0.45155363154015155) (4,0.44003072783936187) (5,0.43598783898569665) (6,0.457939339593458) (7,0.4445345884701096) (8,0.47772974226049786) (9,0.48566555612216905) (10,0.5038136720679858) (11,0.45598904949435654) (12,0.45252314382134246) (13,0.4639174138671322) (14,0.46192813690560464) (15,0.450147942901558) (16,0.4591003975006569) (17,0.45940896533317077) };
            \addlegendentry{UNet (V)}
        \nextgroupplot[
            title={(b) NeWM Comparison},
            legend style={at={(0.95,0.95)}, anchor=north east, legend columns=2, font=\scriptsize},
            legend cell align=left,
            xmin=1,
            xmax=20,
            ymax=1.1
        ]
            \addplot[
                color=barbie_pink,
                mark=o
            ] 
            coordinates {(1,0.7962014808417848) (2,0.643723277016456) (3,0.5834122543564494) (4,0.5497866679237496) (5,0.5270069143416719) (6,0.5091872714135958) (7,0.4938834207768766) (8,0.4813490274900235) (9,0.470823872163429) (10,0.46251336064027704) (11,0.45528372983384574) (12,0.4497386727088727) (13,0.4449051910120508) (14,0.44128113966550886) (15,0.4381608010634132) (16,0.4356967245940096) (17,0.4335165566168957) (18,0.4316443329827386) (19,0.4299434896396554) (20,0.4288183942159511) };
            \addlegendentry{FCNet MLR (T)}
            \addplot[
                color=barbie_pink,
                mark=x
            ] 
            coordinates {(1,0.7158893147505909) (2,0.6295144759028791) (3,0.5856030467583473) (4,0.5577558283525778) (5,0.5361412626253553) (6,0.5179308641208223) (7,0.5025253852447832) (8,0.48970746769603474) (9,0.4792099370654807) (10,0.4703751162412655) (11,0.4631487573665309) (12,0.45734411280557336) (13,0.45241776018975727) (14,0.44841780010835236) (15,0.44498054683208466) (16,0.44239484455930184) (17,0.4399070484810565) (18,0.4379087364278644) };
            \addlegendentry{FCNet MLR (V)}
            %
            \addplot[
                color=plum,
                mark=o
            ] 
            coordinates {(1,0.5358307536529459) (2,0.4716468391396244) (3,0.42616646213931325) (4,0.41064452588187983) (5,0.4009659090397521) (6,0.39204079421780863) (7,0.3852493857374843) (8,0.37911255185648524) (9,0.37435372183041543) (10,0.37096558927749257) (11,0.36854876346469667) (12,0.36702919089646074) (13,0.36509720888567265) };
            \addlegendentry{FCNet ANN (T)}
            \addplot[
                color=plum,
                mark=x
            ] 
            coordinates {(1,0.5060000547982124) (2,0.4513483504394451) (3,0.41791234493076085) (4,0.40725683663264817) (5,0.3988042793360101) (6,0.39177498032888736) (7,0.3857681408344981) (8,0.38034105399645957) (9,0.37869691319135296) (10,0.37479011283581515) (11,0.3742740233260465) (12,0.37335140509418696) (13,0.372078339917114)  };
            \addlegendentry{FCNet ANN (V)}
            %
            \addplot[
                color=jazzberry_jam,
                mark=o
            ] 
            coordinates {(1,1.0282852565064842) (2,0.742680866961126) (3,0.6411708613604675) (4,0.5999477695167801) (5,0.5788145943740268) (6,0.5624155461972142) (7,0.5467426890024433) (8,0.5307433508612491) (9,0.514748517193912) (10,0.49947595108806353) (11,0.4849628197190202) (12,0.47308646032103785) (13,0.4632909075345522) (14,0.4559279416262368) (15,0.4488282389478919) (16,0.44386894522625725) (17,0.44060678798475383) (18,0.43722661924950873) (19,0.4342654308787099) (20,0.4323862984224602) };
            \addlegendentry{PanguW MLR (T)}
            \addplot[
                color=jazzberry_jam,
                mark=x
            ] 
            coordinates {(1,0.8777127081012152) (2,0.7007104618003569) (3,0.6370390280183539) (4,0.6071800866040838) (5,0.5882741662572666) (6,0.571657545505518) (7,0.5552659514977272) (8,0.5389635302575237) (9,0.5227860583957419) (10,0.5070663221090673) (11,0.49322115567075203) (12,0.48167703659778616) (13,0.47191499056945363) (14,0.46400632044996126) (15,0.45736368458314114) (16,0.45218662118696307) (17,0.4479710687714887) (18,0.44449475847453956) (19,0.44149557626750097) (20,0.4390308677432049) };
            \addlegendentry{PanguW MLR (V)}
            %
            \addplot[
                color=dodger_blue,
                mark=o
            ] 
            coordinates {(1,0.5265016231087991) (2,0.44059723762818326) (3,0.4056626508633296) (4,0.3927132620671649) (5,0.380366372289481) (6,0.3727545980134128) (7,0.36698373039195564) (8,0.3629398860129309) (9,0.360499593394774) (10,0.35877886828449035) (11,0.3565240016690007) (12,0.35566972610023284) (13,0.3542612731272792) (14,0.3530058160792162) };
            \addlegendentry{PanguW ANN (T)}
            \addplot[
                color=dodger_blue,
                mark=x
            ] 
            coordinates {(1,0.4955298553209707) (2,0.41544645333505537) (3,0.40094357486589843) (4,0.38677140854927433) (5,0.37775698250316714) (6,0.3739656871521329) (7,0.36921661679285117) (8,0.36743274052818137) (9,0.3619588200586388) (10,0.3605978534702795) (11,0.3594074409948774) (12,0.3585718707686447) (13,0.3566134505602251) (14,0.35634245032287504) };
            \addlegendentry{Pangu ANN (V)}
        \end{groupplot}
    \end{tikzpicture}
    
    \caption{Training curves for (a) comparing across algorithms trained on PanguWeather outputs and (b) comparing algorithm performance when trained on PanguWeather (P) and FourCastNet v2 (F) outputs. All inputs were masked, T denotes Training, V denotes Validation.}
    \label{fig:training_curves}
\end{figure*}

\begin{table*}[h]
    \centering
    \begin{tabular}{ccrcl|rcl}
          &Postprocessing&  \multicolumn{3}{c}{RMSE}&  \multicolumn{3}{c}{CRPS}\\
          NeWM&Algorithm&  Training&  Validation&  Test&  Training&  Validation& Test\\
          \hline
        \multirow{4}{*}{\centering\shortstack{PanguWeather\\(Masked)}}&MLR&  \num{0.7226505182039591}&  \num{0.8063037991523743}&  \num{0.6480445861816406}&  \num{0.4323862984224602}&  \num{0.4390308677432049} & \num{0.4448533356189728}\\
             & ANN&  \num{0.5230642033763874}&  \textbf{\num{0.5577106475830078}}&  \textbf{\num{0.44764015078544617}}&  \num{0.3530058160792162}&  \textbf{\num{0.3609984517097473}} & \textbf{\num{0.31441423296928406}}\\
             & CNN&  \textit{\num{0.12174003649551642}}&  \num{0.6346365213394165}&  \num{0.5550320744514465}&  \textbf{\num{0.2253519843248072}}&  \num{0.43800976872444153} & \num{0.37968122959136963}\\
             & UNet&  \num{0.22527445304650867}&  \num{0.5700232982635498}&  \num{0.5116170048713684}&  \num{0.28434710794279217}&  \num{0.3657682240009308} & \num{0.33425313234329224}\\
             &&&&&&&\\
        \multirow{4}{*}{\centering\shortstack{PanguWeather\\(Unmasked)}}&MLR&  \num{0.8327379979101228}&  \num{0.8650394082069397}&  \num{0.6751123070716858}&  \num{0.4288718418022733}&  \num{0.43326464717287616} & \num{0.44098952412605286}\\
             & ANN&  \num{0.6874208384089999}&  \num{0.6939778327941895}&  \textit{\num{0.5672893524169922}}&  \num{0.36712426830221107}&  \textit{\num{0.3763017952442169}} & \textit{\num{0.3342660069465637}}\\
             & CNN& \textbf{\num{0.09049266133550296}}&  \num{0.682407021522522}&  \num{0.6006266474723816}&  \num{0.4258285040823801}&  \num{0.45354679226875305} & \num{0.3956443965435028}\\
             & UNet&  \num{0.32420794769007083}&  \textit{\num{0.6474366188049316}}&  \textit{\num{0.5702997446060181}}&  \textit{\num{0.3110308501505314}}&  \num{0.3999064564704895} & \num{0.35435590147972107}\\
             &&&&&&&\\
        \multirow{4}{*}{\centering\shortstack{FourCastNet v2\\(Masked)}}&MLR&  \num{0.7233120131751766}&  \num{0.760333538055419}&  \num{0.56929874420166}&  \num{0.4288183942159511}&  \num{0.440880209207534} & \num{0.37368956208229}\\
             & ANN&  \num{0.5861701613627606}&  \num{0.614731967449188}&  \textit{\num{0.48458844423294}}&  \num{0.36509720888567265}&  \textit{\num{0.376264184713363}} & \textit{\num{0.32447811961174}}\\
             & CNN&  \textit{\num{0.12499715748634545}}&  \num{0.64563912153244}&  \num{0.554476141929626}&  \textbf{\num{0.23008192367377175}}&  \num{0.4299445534132391} & \num{0.372331231832504}\\
             & UNet&  \num{0.259527009401684}&  \textbf{\num{0.556589245796203}}&  \num{0.49734815955162}&  \num{0.26829113579906855}&  \num{0.390459835529327} & \num{0.354544311761856}\\
             &&&&&&&\\
        \multirow{4}{*}{\centering\shortstack{FourCastNet v2\\(Unmasked)}}&MLR&  \num{0.7880482205322811}&  \num{0.815926432609558}&  \num{0.64768225}&  \num{0.4518836188760603}&  \num{0.464443981647491} & \num{0.400884211}\\
             & ANN&  \num{0.6136063043004978}&  \textit{\num{0.652801752090454}}&  \textit{\num{0.531455159187316}}&  \num{0.37466548318448273}&  \textit{\num{0.388157695531845}} & \textit{\num{0.339153498411178}}\\
             & CNN&  \textit{\num{0.09424427436563856}}&  \num{0.700313150882721}&  \num{0.5994154214859}&  \num{0.3339094364256331}&  \num{0.457492798566818} & \num{0.405330300331115}\\
             & UNet&  \num{0.40494587991659525}&  \num{0.67156833410263}&  \num{0.584327399730682}&  \textit{\num{0.30565360093572336}}&  \num{0.420835644006729} & \num{0.369675278663635}\\
             &&&&&&&\\
        
    \end{tabular}
    \caption{Deterministic and Probabilistic overall normalized root-mean squared error (RMSE) and normalized continuous ranked probability score (CRPS) for PanguWeather and FourCastNetv2 using masked and unmasked inputs. Bold values indicate set minima, while italicized values indicate per-input source set minima.}
    \label{tab:results}
\end{table*}
We begin by evaluating the impact of the choice of algorithm on postprocessing performance when trained on PanguWeather outputs and the impact on algorithm performance when trained on PanguWeather vs when trained on FourCastNet v2. We do this by first comparing the curves measuring the CRPS as a function of training epoch, which are shown in ~\Cref{fig:training_curves}. We note here that as expected increasing model expressivity results in lower CRPS values when comparing performance on the training set, indicating that more expressive models are able to learn patterns in the data that allow them to fit the training data. However, we see that the models we train using the 2D fields directly have a problem generalizing from the training to validation set. 
\par We found this surprising, given that the pixel-wide distributions for the fields were found to be a reasonable match (see \Cref{fig:distSample}) and that the MLR and ANN models generalize well between sets. We proceeded to attempt a number of strategies for addressing this overfitting behavior, including batch normalization, dropout layers, and L2 regularization (i.e., weight decay). More details for our regularization strategies are given in Section S-4,  
but none of the strategies resolved the generalization issue experienced by the proposed CNN architectures. 
\par Next, we note that the choice of NeWM (among those we test) does not significantly impact the performance of the post-processing algorithm, as evidenced both by the training curves and \Cref{tab:results}. This suggests that the representation of the atmosphere by the NeWMs is comparable for the purposes of field postprocessing.

\tikzsetnextfilename{Figure_5}

\subsection{Linear Models}
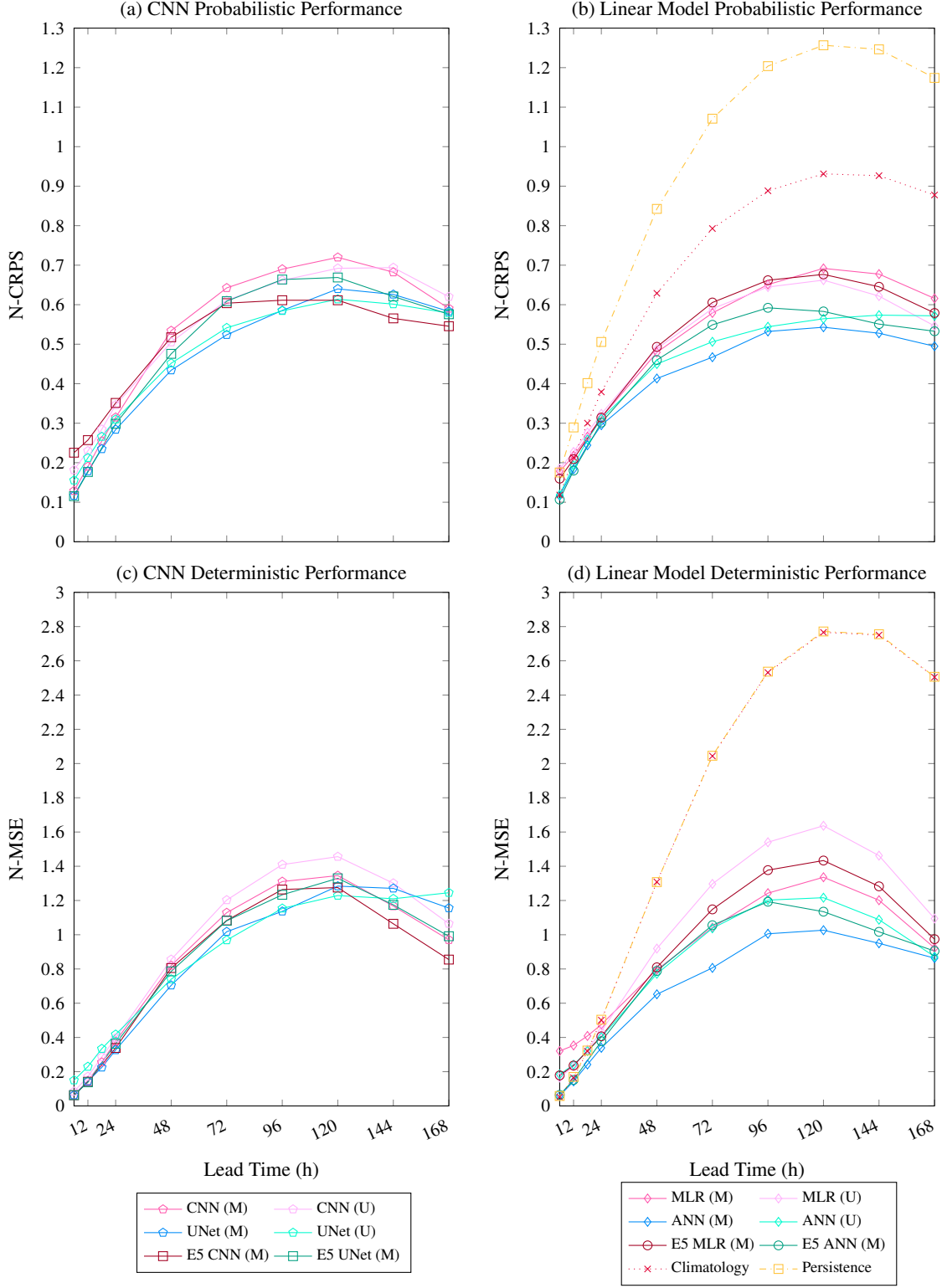
\begin{figure*}[h]
    \centering
    \begin{tikzpicture}
        \begin{groupplot}[
            group style={
                group name=Per Lead Time Curves,
                group size=2 by 2,
                x descriptions at=edge bottom,
                vertical sep=.8cm,
                horizontal sep = 1.75cm
            },
            footnotesize,
            width=7.5cm,
            height=9.7cm,
            xlabel=Lead Time (h),
        ]
        \nextgroupplot[
            title={(a) CNN Probabilistic Performance},
            legend cell align=left,
            legend style={font=\scriptsize},
            legend style={at={(.9,-0.3)}, anchor=south east, legend columns=2},
            xmin=6,
            xmax=168,
            ymax=1.3,
            ymin=0,
            ylabel= N-CRPS ,
            xtick={12, 24, 48, 72, 96, 120, 144, 168},
            xticklabel style={rotate=25, anchor=north east},
        ]
        \addplot[
        color=barbie_pink,
                mark=pentagon
        ] 
        table [x=Leadtime, y=m_cnn_p, col sep=comma]{csvs/model_comparison_leadtime_pangu_test-reval.csv};

        \addplot[
        color=plum,
                mark=pentagon
        ] 
        table [x=Leadtime, y=u_cnn_p, col sep=comma]{csvs/model_comparison_leadtime_pangu_test-reval.csv};

        \addplot[
        color=dodger_blue,
                mark=pentagon
        ] 
        table [x=Leadtime, y=m_unet_p, col sep=comma]{csvs/model_comparison_leadtime_pangu_test-reval.csv};

        \addplot[
        color=aquamarine,
                mark=pentagon
        ] 
        table [x=Leadtime, y=u_unet_p, col sep=comma]{csvs/model_comparison_leadtime_pangu_test-reval.csv};

        \addplot[
        color=carmine,
                mark=square
        ] 
        table [x=Leadtime, y=m_cnn_p, col sep=comma]{csvs/model_comparison_leadtime_ERA5_test-reval.csv};
                \addplot[
        color=jazzberry_creepers,
                mark=square
        ] 
        table [x=Leadtime, y=m_unet_p, col sep=comma]{csvs/model_comparison_leadtime_ERA5_test-reval.csv};
        
        \nextgroupplot[
            title={(b) Linear Model Probabilistic Performance},
            legend cell align=left,
            legend style={font=\scriptsize},
            legend style={at={(.9,-0.3)}, anchor=south east, legend columns=2},
            xmin=6,
            xmax=168,
            ymax=1.3,
            ymin=0,
            ylabel= N-CRPS,
            xtick={12, 24, 48, 72, 96, 120, 144, 168},
            xticklabel style={rotate=25, anchor=north east},
        ]
        \addplot[
        color=barbie_pink,
                mark=diamond
        ] 
        table [x=Leadtime, y=m_mlr_p, col sep=comma]{csvs/model_comparison_leadtime_pangu_test-reval.csv};

        \addplot[
        color=plum,
                mark=diamond
        ] 
        table [x=Leadtime, y=u_mlr_p, col sep=comma]{csvs/model_comparison_leadtime_pangu_test-reval.csv};

        \addplot[
        color=dodger_blue,
                mark=diamond
        ] 
        table [x=Leadtime, y=m_ann_p, col sep=comma]{csvs/model_comparison_leadtime_pangu_test-reval.csv};

        \addplot[
        color=aquamarine,
                mark=diamond
        ] 
        table [x=Leadtime, y=u_ann_p, col sep=comma]{csvs/model_comparison_leadtime_pangu_test-reval.csv};

        \addplot[
        color=carmine,
                mark=o
        ] 
        table [x=Leadtime, y=m_mlr_p, col sep=comma]{csvs/model_comparison_leadtime_ERA5_test-reval.csv};
                \addplot[
        color=jazzberry_creepers,
                mark=o
        ] 
        table [x=Leadtime, y=m_ann_p, col sep=comma]{csvs/model_comparison_leadtime_ERA5_test-reval.csv};
        \addplot[
        color = alizarin_crimson,
            dotted,
            mark=x,
            mark options={style=solid}
        ]
        table [x=Leadtime, y=clim_p, col sep=comma]{csvs/model_comparison_baselines.csv};
        \addplot[
        color = bright_spark,
            dashdotted,
            mark=square,
            mark options={style=solid}
        ]
        table [x=Leadtime, y=pers_p, col sep=comma]{csvs/model_comparison_baselines.csv};

        \nextgroupplot[
            title={(c) CNN Deterministic Performance},
            legend cell align=left,
            legend style={font=\scriptsize},
            legend style={at={(.9,-0.325)}, anchor=south east, legend columns=2},
            xmin=6,
            xmax=168,
            ymax=3,
            ymin=0,
            ylabel= N-MSE,
            xtick={12, 24, 48, 72, 96, 120, 144, 168},
            xticklabel style={rotate=25, anchor=north east},
        ]
        \addplot[
        color=barbie_pink,
                mark=pentagon
        ] 
        table [x=Leadtime, y=m_cnn_d, col sep=comma]{csvs/model_comparison_leadtime_pangu_test-reval.csv};
        \addlegendentry{CNN (M)}

        \addplot[
        color=plum,
                mark=pentagon
        ] 
        table [x=Leadtime, y=u_cnn_d, col sep=comma]{csvs/model_comparison_leadtime_pangu_test-reval.csv};
        \addlegendentry{CNN (U)}

        \addplot[
        color=dodger_blue,
                mark=pentagon
        ] 
        table [x=Leadtime, y=m_unet_d, col sep=comma]{csvs/model_comparison_leadtime_pangu_test-reval.csv};
        \addlegendentry{UNet (M)}

        \addplot[
        color=aquamarine,
                mark=pentagon
        ] 
        table [x=Leadtime, y=u_unet_d, col sep=comma]{csvs/model_comparison_leadtime_pangu_test-reval.csv};
        \addlegendentry{UNet (U)}

        \addplot[
        color=carmine,
                mark=square
        ] 
        table [x=Leadtime, y=m_cnn_d, col sep=comma]{csvs/model_comparison_leadtime_ERA5_test-reval.csv};
        \addlegendentry{E5 CNN (M)}
                \addplot[
        color=jazzberry_creepers,
                mark=square
        ] 
        table [x=Leadtime, y=m_unet_d, col sep=comma]{csvs/model_comparison_leadtime_ERA5_test-reval.csv};
        \addlegendentry{E5 UNet (M)}

        \nextgroupplot[
            title={(d) Linear Model Deterministic Performance},
            legend cell align=left,
            legend style={font=\scriptsize},
            legend style={at={(.9,-0.35)}, anchor=south east, legend columns=2},
            xmin=6,
            xmax=168,
            ymax=3,
            ymin=0,
            ylabel=N-MSE,
            xtick={12, 24, 48, 72, 96, 120, 144, 168},
            xticklabel style={rotate=25, anchor=north east},
        ]
        \addplot[
        color=barbie_pink,
                mark=diamond
        ] 
        table [x=Leadtime, y=m_mlr_d, col sep=comma]{csvs/model_comparison_leadtime_pangu_test-reval.csv};
        \addlegendentry{MLR (M)}

        \addplot[
        color=plum,
                mark=diamond
        ] 
        table [x=Leadtime, y=u_mlr_d, col sep=comma]{csvs/model_comparison_leadtime_pangu_test-reval.csv};
        \addlegendentry{MLR (U)}

        \addplot[
        color=dodger_blue,
                mark=diamond
        ] 
        table [x=Leadtime, y=m_ann_d, col sep=comma]{csvs/model_comparison_leadtime_pangu_test-reval.csv};
        \addlegendentry{ANN (M)}

        \addplot[
        color=aquamarine,
                mark=diamond
        ] 
        table [x=Leadtime, y=u_ann_d, col sep=comma]{csvs/model_comparison_leadtime_pangu_test-reval.csv};
        \addlegendentry{ANN (U)}

        \addplot[
        color=carmine,
                mark=o
        ] 
        table [x=Leadtime, y=m_mlr_d, col sep=comma]{csvs/model_comparison_leadtime_ERA5_test-reval.csv};
        \addlegendentry{E5 MLR (M)}
                \addplot[
        color=jazzberry_creepers,
                mark=o
        ] 
        table [x=Leadtime, y=m_ann_d, col sep=comma]{csvs/model_comparison_leadtime_ERA5_test-reval.csv};
        \addlegendentry{E5 ANN (M)}
        \addplot[
        color = alizarin_crimson,
            dotted,
            mark=x,
            mark options={style=solid}
        ]
        table [x=Leadtime, y=clim_d, col sep=comma]{csvs/model_comparison_baselines.csv};
        \addlegendentry{Climatology}
        \addplot[
        color = bright_spark,
            dashdotted,
            mark=square,
            mark options={style=solid}
        ]
        table [x=Leadtime, y=pers_d, col sep=comma]{csvs/model_comparison_baselines.csv};
        \addlegendentry{Persistence}
        \end{groupplot}
    \end{tikzpicture}
    \caption{Breakdown of CNN model performance across lead times on the test set, evaluated probabilistically for (a) CNNs and (b) linear models, as well as evaluated deterministically for (c) CNNs and (d) linear models. In (b), the persistence baseline performance is computed using MAE.}
    \label{fig:leadtime_breakdown}
\end{figure*}

We now focus on the linear models trained on both masked and unmasked inputs. In \Cref{fig:leadtime_breakdown}, we show that performance between masked and unmasked models is quite similar, though we find a small but consistent improvement when using masked inputs. Furthermore, the inherent interpretability of linear models allows us to direcly look at the model weights, shown in \Cref{tab:linCoefficientsTransposed} to interpret model behavior. Here we point to $\alpha_7$, which corresponds to the lead time input feature for the linear models. For the unmasked inputs--featured at the bottom-- the probabilistic linear model learns an anti-correlation between the lead time and the standard deviation of the change in pressure. 
\par As we know that in general there is an increase in the uncertainty of our prediction at longer lead times (as there generally is greater variability in the observed intensification at longer lead times) , this learned anti-correlation suggests that the model learns to compensate for the variability in other features through the lead time feature instead of learning a direct relationship. This behavior is not perceived reflected in the standard deviation coefficients for the models trained on masked inputs. We also see that the masked model in general relies on more varied sources of NeWM data (as evidenced by coefficients \makebox{$\alpha_1$ - $\alpha_6$}).
\par
\begin{figure}[h]
    \centering
    \begin{subfigure}[]{0.3\textwidth}
        \centering
        \includegraphics[width=.95\textwidth]{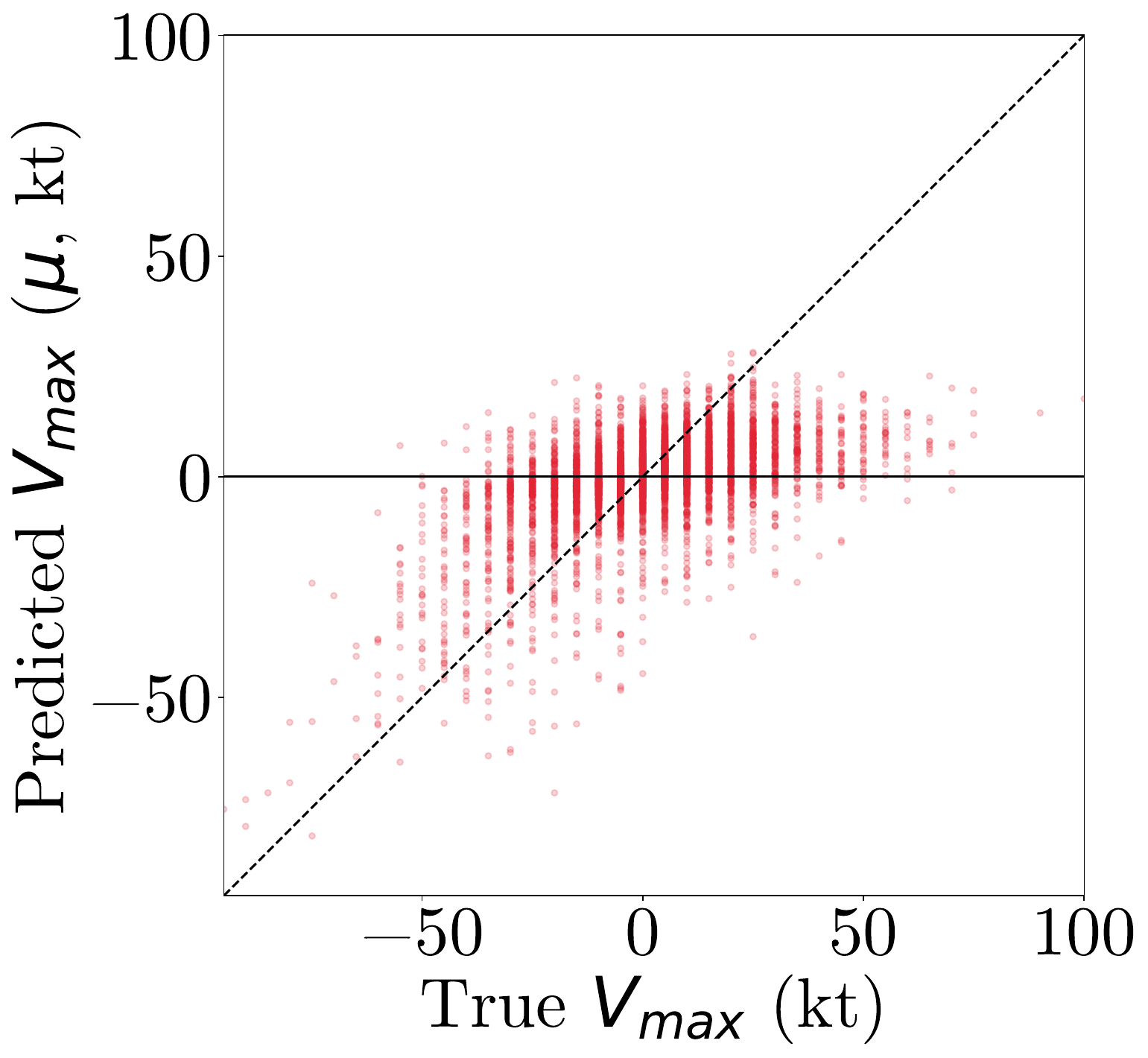}
        \caption{Lead time 24h}
        \label{fig:24_pred_v_true}
    \end{subfigure}
    \hfill\\
    \begin{subfigure}[]{0.3\textwidth}
        \centering
        \includegraphics[width=.95\textwidth]{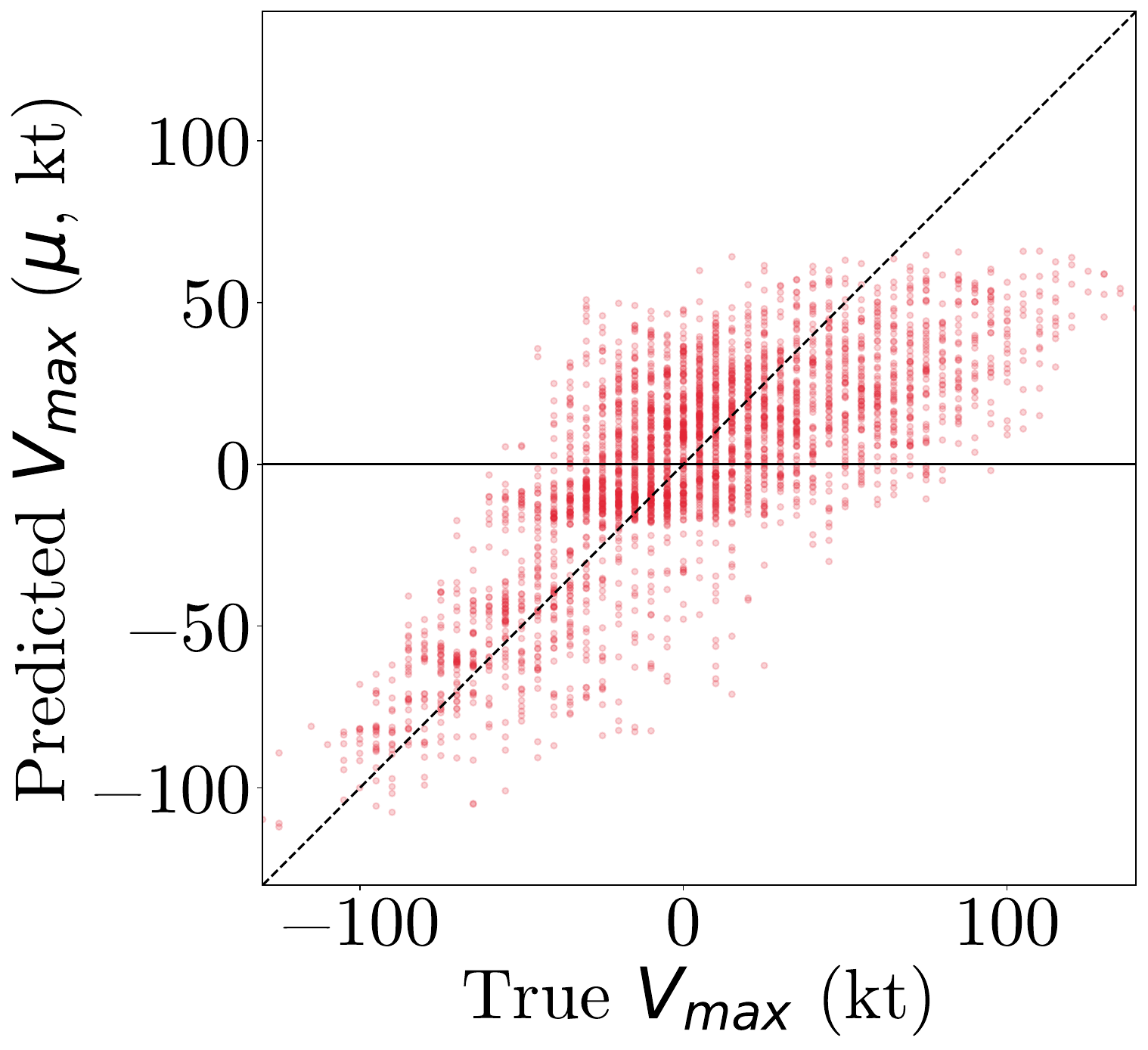}
        \caption{Lead time 96h}
        \label{fig:72_pred_v_true}
    \end{subfigure}\\
    \begin{subfigure}[]{0.3\textwidth}
        \centering
        \includegraphics[width=.95\textwidth]{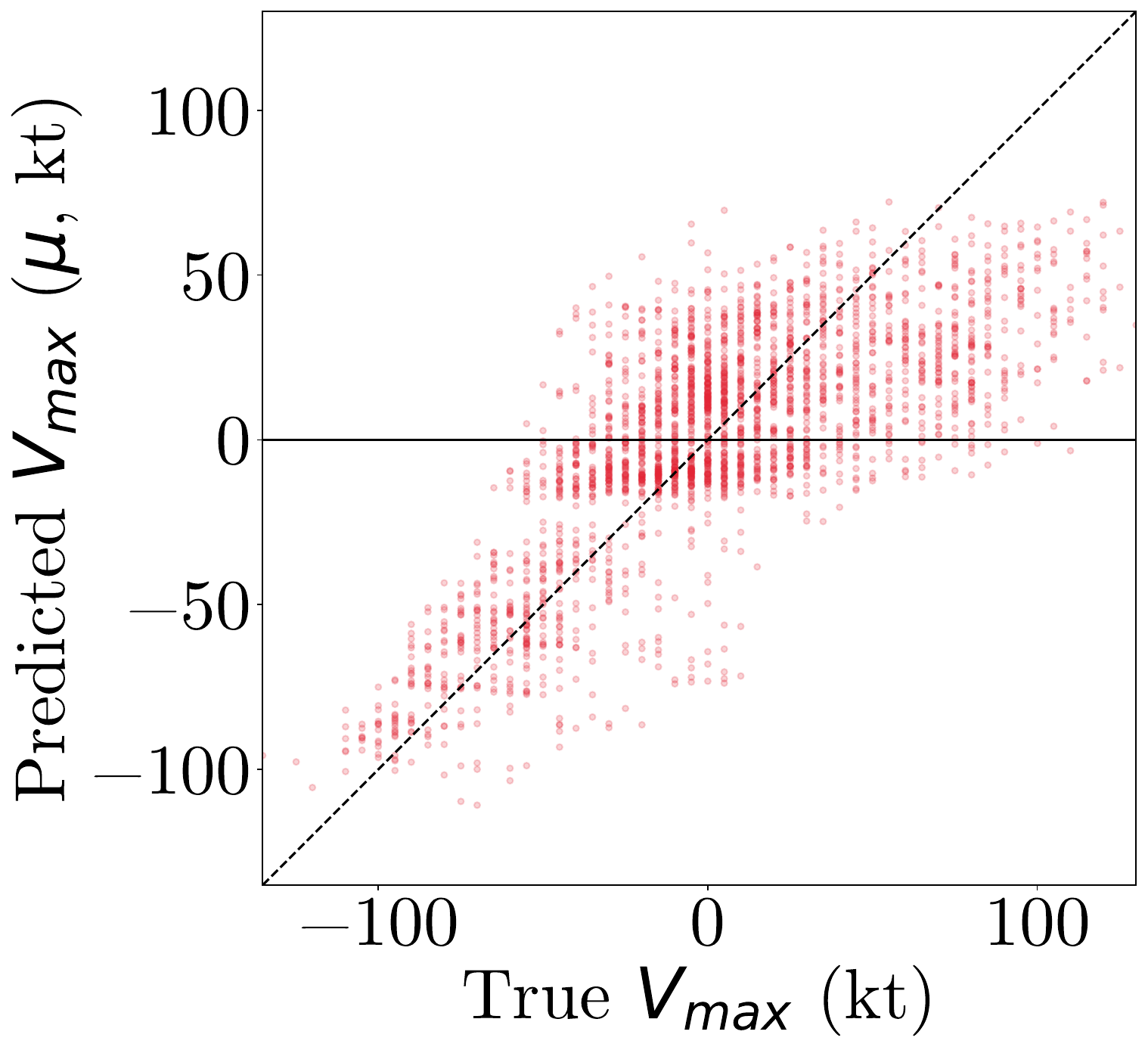}
        \caption{Lead time 120h}
        \label{fig:120_pred_v_true}
    \end{subfigure}
    \caption{Predicted Mean Intensification vs True Intensification for all test set samples. Point alpha is set to 0.2 for each sample.}
    \label{fig:pred_v_true}
\end{figure}

 We also provide plots showing the mean intensification ($\mu_{\Delta V_{max}}$) predicted by the ANN vs. the observed intensification from IBTrACS for lead times $\tau \in \left\{ 24, 96, 120 \right\}$ in \Cref{fig:pred_v_true}. We choose $\tau = 24h$ as it is the point where the linear models' probabilistic performance is most similar per (b) in \Cref{fig:leadtime_breakdown}, $\tau = 96h$ as it is the point at which the last of the deterministic models have an error of about 1 standard deviation per (d) in \Cref{fig:leadtime_breakdown}, and $\tau = 120h$ as it is the maximum horizon reported by some operational products (e.g., the 5-day NHC forecast). In \Cref{fig:pred_v_true} we see that the predicted mean intensification generally fall below the y=x line for the higher intensification rates, while the de-intensification tail is better captured. This further emphasizes the benefit of predicting a full distribution per sample, as the predicted variance can help capture the probability of the event happening. 

\begin{table}[h]
    \centering
    \begin{tabular}{crrrr}
         & \multicolumn{4}{c}{PanguWeather (Masked)} \\
         Coefficient & $\mu_{\Delta V_{max}}$ & $\sigma_{\Delta V_{max}}$ & $\mu_{\Delta P_{min}}$ & $\sigma_{\Delta P_{min}}$ \\
         \hline
        $\alpha_1$ & \num{0.2429} & \num{0.1036} & \num{-0.0921} & \num{0.0848} \\
        $\alpha_2$ & \num{-0.1642} & \num{-0.0555} & \num{-0.0299} & \num{0.1195} \\
        $\alpha_3$ & \num{-0.0525} & \num{-0.0101} & \num{-0.0135} & \num{-0.0125} \\
        $\alpha_4$ & \num{-0.1646} & \num{-0.0368} & \num{-0.1143} & \num{0.1784} \\
        $\alpha_5$ & \num{0.0423} & \num{0.1121} & \num{-0.0207} & \num{0.0617} \\
        $\alpha_6$ & \num{-0.1029} & \num{0.1405} & \num{0.1679} & \num{0.0237} \\
        $\alpha_7$ & \num{0.816} & \num{0.7541} & \num{-0.2651} & \num{0.4767} \\
        $\alpha_8$ & \num{-0.3399} & \num{0.0913} & \num{0.3293} & \num{0.0161} \\
        $\alpha_9$ & \num{0.3961} & \num{-0.0861} & \num{-0.375} & \num{-0.1965} \\
        \\
        & \multicolumn{4}{c}{PanguWeather (Unmasked)} \\
        Coefficient & $\mu_{\Delta V_{max}}$ & $\sigma_{\Delta V_{max}}$ & $\mu_{\Delta P_{min}}$ & $\sigma_{\Delta P_{min}}$ \\
        \hline
        $\alpha_1$ & \num{0.0546} & \num{-0.0475} & \num{-0.0092} & \num{-0.1009} \\
        $\alpha_2$ & \num{-0.0038} & \num{-0.0206} & \num{0.0705} & \num{-0.0261} \\
        $\alpha_3$ & \num{0.1173} & \num{0.1239} & \num{-0.1377} & \num{-0.0115} \\
        $\alpha_4$ & \num{-0.0101} & \num{-0.0251} & \num{0.0619} & \num{-0.0183} \\
        $\alpha_5$ & \num{0.0621} & \num{0.0254} & \num{-0.0646} & \num{-0.0116} \\
        $\alpha_6$ & \num{-0.0335} & \num{0.01} & \num{0.0231} & \num{0.0019} \\
        $\alpha_7$ & \num{0.0665} & \num{1.0301} & \num{-0.1335} & \num{-0.7796} \\
        $\alpha_8$ & \num{-0.3035} & \num{0.1347} & \num{0.2858} & \num{-0.0994} \\
        $\alpha_9$ & \num{0.3259} & \num{-0.0894} & \num{-0.3432} & \num{0.1462} \\
         \\
    \end{tabular}
    \caption{Linear model coefficients for probabilistic MLRs trained on PanguWeather. $\alpha_0$ - $\alpha_8$ corresponding row-wise to $\mathbf{x}_{lin}$ in \Cref{eq:linInputs} as described by \Cref{eq:linModel}}
    \label{tab:linCoefficientsTransposed}
\end{table}

\tikzsetnextfilename{Figure_6}

\subsection{Case Study - Claudia 2020}
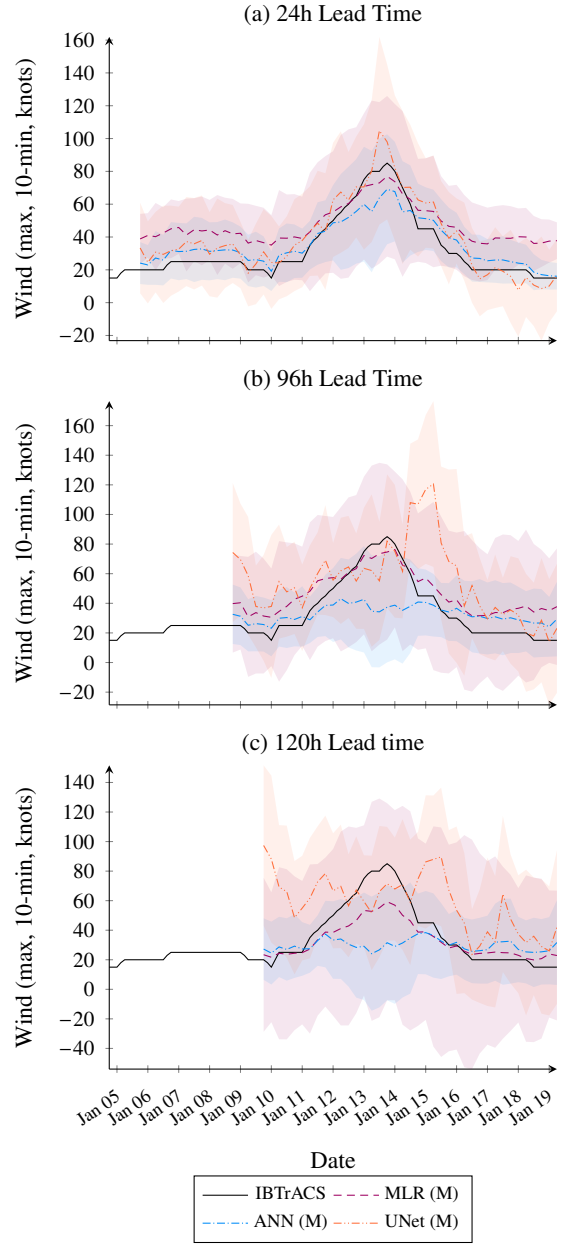
\begin{figure}[h]
    \centering
    \begin{tikzpicture}
        \begin{groupplot}[
            group style={
                group size=1 by 3,
                x descriptions at=edge bottom,
                vertical sep=0.8cm,
            },
            footnotesize,
            width=7.5cm,
            height=5.6cm,
            date coordinates in=x,
            xticklabel={
                \pgfcalendarmonthshortname{\month} \day
            },
            xticklabel style={font=\scriptsize},
            xlabel={Date},
            xticklabel style={rotate=35, anchor=north east},
            legend columns=2,
        ]
        \nextgroupplot[
            title={(a) 24h Lead Time},
            legend style={font=\scriptsize, at={(0.5,-0.07)}, anchor=north},
            axis lines=left,
            ylabel={Wind (max, 10-min, knots)},
        ]
        \addplot[
        ] 
        table [x=time, y=max_wind, col sep=comma]{csvs/CLAUDIA_IBTrACS.csv};
        \addplot[
            color=jazzberry_jam,
            densely dashed,
            name path=mlr_meanw_18,
        ] 
        table [x=time, y=mean_max_wind, col sep=comma]{csvs/CLAUDIA_MLR (Masked)_24_hour_forecast.csv};
        \addplot[
            color=dodger_blue,
            densely dashdotted,
            name path=ann_meanw_18,
        ] 
        table [x=time, y=mean_max_wind, col sep=comma]{csvs/CLAUDIA_ANN (LeakyReLU, M)_24_hour_forecast.csv};
        \addplot[
            color=outrageous_orange,
            densely dashdotdotted,
            name path=unet_meanw_18,
        ] 
        table [x=time, y=mean_max_wind, col sep=comma]{csvs/CLAUDIA_UNetv2 (dout 0.33)_24_hour_forecast.csv};
        \addplot[
            name path=mlr_upperw_18,
            draw=none,
        ] 
        table [x=time, y=upper_max_wind, col sep=comma]{csvs/CLAUDIA_MLR (Masked)_24_hour_forecast.csv};
        \addplot[
            name path=mlr_lowerw_18,
            draw=none,
        ] 
        table [x=time, y=lower_max_wind, col sep=comma]{csvs/CLAUDIA_MLR (Masked)_24_hour_forecast.csv};
        \addplot [
            thick,
            color=jazzberry_jam,
            fill=jazzberry_jam,
            fill opacity=0.1
        ] fill between[
            of=mlr_upperw_18 and mlr_lowerw_18
        ];
        \addplot[
            name path=ann_upperw_18,
            draw=none,
        ] 
        table [x=time, y=upper_max_wind, col sep=comma]{csvs/CLAUDIA_ANN (LeakyReLU, M)_24_hour_forecast.csv};
        \addplot[
            name path=ann_lowerw_18,
            draw=none,
        ] 
        table [x=time, y=lower_max_wind, col sep=comma]{csvs/CLAUDIA_ANN (LeakyReLU, M)_24_hour_forecast.csv};
        \addplot [
            thick,
            color=dodger_blue,
            fill=dodger_blue,
            fill opacity=0.1
        ] fill between[
            of=ann_upperw_18 and ann_lowerw_18
        ];
        \addplot[
            name path=unet_upperw_18,
            draw=none,
        ] 
        table [x=time, y=upper_max_wind, col sep=comma]{csvs/CLAUDIA_UNetv2 (dout 0.33)_24_hour_forecast.csv};
        \addplot[
            name path=unet_lowerw_18,
            draw=none,
        ] 
        table [x=time, y=lower_max_wind, col sep=comma]{csvs/CLAUDIA_UNetv2 (dout 0.33)_24_hour_forecast.csv};
        \addplot [
            thick,
            color=outrageous_orange,
            fill=outrageous_orange,
            fill opacity=0.1
        ] fill between[
            of=unet_upperw_18 and unet_lowerw_18
        ];
        \nextgroupplot[
            title={(b) 96h Lead Time},
            legend style={font=\scriptsize, at={(0.5,-0.07)}, anchor=north},
            axis lines=left,
            ylabel={Wind (max, 10-min, knots)},
        ]
        \addplot[
        ] 
        table [x=time, y=max_wind, col sep=comma]{csvs/CLAUDIA_IBTrACS.csv};

        \addplot[
            color=jazzberry_jam,
            densely dashed,
            name path=mlr_meanw_72,
        ] 
        table [x=time, y=mean_max_wind, col sep=comma]{csvs/CLAUDIA_MLR (Masked)_96_hour_forecast.csv};
        \addplot[
            color=dodger_blue,
            densely dashdotted,
            name path=ann_meanw_72,
        ] 
        table [x=time, y=mean_max_wind, col sep=comma]{csvs/CLAUDIA_ANN (LeakyReLU, M)_96_hour_forecast.csv};
        \addplot[
            color=outrageous_orange,
            densely dashdotdotted,
            name path=unet_meanw_72,
        ] 
        table [x=time, y=mean_max_wind, col sep=comma]{csvs/CLAUDIA_UNetv2 (dout 0.33)_96_hour_forecast.csv};

        \addplot[
            name path=mlr_upperw_72,
            draw=none,
        ] 
        table [x=time, y=upper_max_wind, col sep=comma]{csvs/CLAUDIA_MLR (Masked)_96_hour_forecast.csv};
        \addplot[
            name path=mlr_lowerw_72,
            draw=none,
        ] 
        table [x=time, y=lower_max_wind, col sep=comma]{csvs/CLAUDIA_MLR (Masked)_96_hour_forecast.csv};
        \addplot [
            thick,
            color=jazzberry_jam,
            fill=jazzberry_jam,
            fill opacity=0.1
        ] fill between[
            of=mlr_upperw_72 and mlr_lowerw_72
        ];
        \addplot[
            name path=ann_upperw_72,
            draw=none,
        ] 
        table [x=time, y=upper_max_wind, col sep=comma]{csvs/CLAUDIA_ANN (LeakyReLU, M)_96_hour_forecast.csv};
        \addplot[
            name path=ann_lowerw_72,
            draw=none,
        ] 
        table [x=time, y=lower_max_wind, col sep=comma]{csvs/CLAUDIA_ANN (LeakyReLU, M)_96_hour_forecast.csv};
        \addplot [
            thick,
            color=dodger_blue,
            fill=dodger_blue,
            fill opacity=0.1
        ] fill between[
            of=ann_upperw_72 and ann_lowerw_72
        ];
        \addplot[
            name path=unet_upperw_72,
            draw=none,
        ] 
        table [x=time, y=upper_max_wind, col sep=comma]{csvs/CLAUDIA_UNetv2 (dout 0.33)_96_hour_forecast.csv};
        \addplot[
            name path=unet_lowerw_72,
            draw=none,
        ] 
        table [x=time, y=lower_max_wind, col sep=comma]{csvs/CLAUDIA_UNetv2 (dout 0.33)_96_hour_forecast.csv};
        \addplot [
            thick,
            color=outrageous_orange,
            fill=outrageous_orange,
            fill opacity=0.1
        ] fill between[
            of=unet_upperw_72 and unet_lowerw_72
        ];
        \nextgroupplot[
            title={(c) 120h Lead time},
            legend style={font=\scriptsize, at={(0.5,-0.35)}, anchor=north},
            axis lines=left,
            ylabel={Wind (max, 10-min, knots)},
        ]
        \addplot[
        ] 
        table [x=time, y=max_wind, col sep=comma]{csvs/CLAUDIA_IBTrACS.csv};
        \addlegendentry{IBTrACS}

        \addplot[
            color=jazzberry_jam,
            densely dashed,
            name path=mlr_meanw_72,
        ] 
        table [x=time, y=mean_max_wind, col sep=comma]{csvs/CLAUDIA_masked MLR_120_hour_forecast.csv};
        \addlegendentry{MLR (M)}
        \addplot[
            color=dodger_blue,
            densely dashdotted,
            name path=ann_meanw_120,
        ] 
        table [x=time, y=mean_max_wind, col sep=comma]{csvs/CLAUDIA_ANN (LeakyReLU, M)_120_hour_forecast.csv};
        \addlegendentry{ANN (M)}
        \addplot[
            color=outrageous_orange,
            densely dashdotdotted,
            name path=unet_meanw_120,
        ] 
        table [x=time, y=mean_max_wind, col sep=comma]{csvs/CLAUDIA_UNetv2 (dout 0.33)_120_hour_forecast.csv};
        \addlegendentry{UNet (M)}
        \addplot[
            name path=mlr_upperw_120,
            draw=none,
        ] 
        table [x=time, y=upper_max_wind, col sep=comma]{csvs/CLAUDIA_masked MLR_120_hour_forecast.csv};
        \addplot[
            name path=mlr_lowerw_120,
            draw=none,
        ] 
        table [x=time, y=lower_max_wind, col sep=comma]{csvs/CLAUDIA_masked MLR_120_hour_forecast.csv};
        \addplot [
            thick,
            color=jazzberry_jam,
            fill=jazzberry_jam,
            fill opacity=0.1
        ] fill between[
            of=mlr_upperw_120 and mlr_lowerw_120
        ];
        \addplot[
            name path=ann_upperw_120,
            draw=none,
        ] 
        table [x=time, y=upper_max_wind, col sep=comma]{csvs/CLAUDIA_ANN (LeakyReLU, M)_120_hour_forecast.csv};
        \addplot[
            name path=ann_lowerw_120,
            draw=none,
        ] 
        table [x=time, y=lower_max_wind, col sep=comma]{csvs/CLAUDIA_ANN (LeakyReLU, M)_120_hour_forecast.csv};
        \addplot [
            thick,
            color=dodger_blue,
            fill=dodger_blue,
            fill opacity=0.1
        ] fill between[
            of=ann_upperw_120 and ann_lowerw_120
        ];
        \addplot[
            name path=unet_upperw_120,
            draw=none,
        ] 
        table [x=time, y=upper_max_wind, col sep=comma]{csvs/CLAUDIA_UNetv2 (dout 0.33)_120_hour_forecast.csv};
        \addplot[
            name path=unet_lowerw_120,
            draw=none,
        ] 
        table [x=time, y=lower_max_wind, col sep=comma]{csvs/CLAUDIA_UNetv2 (dout 0.33)_120_hour_forecast.csv};
        \addplot [
            thick,
            color=outrageous_orange,
            fill=outrageous_orange,
            fill opacity=0.1
        ] fill between[
            of=unet_upperw_120 and unet_lowerw_120
        ];
        \end{groupplot}
    \end{tikzpicture}
    \caption{(a) 24h, (b) 96h, and (c) 120h forecast curves for maximum sustained wind speeds for TC Claudia (2020)}
    \label{fig:claudiastudy}
\end{figure}

In this section we discuss the postprocessing model predictions for Severe Tropical Cyclone Claudia, which formed in the Australian basin  in January 2020. We select it due both its presence in the test set, and the length of the event (which allows a better look at how the postprocessing models perform at longer lead times). While we only discuss the predictions for the Maximum Wind time series (shown in \Cref{fig:claudiastudy}) in this section, both maximum wind and minimum pressure predictions are presented in Section S-6, where we also provide prediction graphs for the most intense storm of the test set for each basin and a second, less intense storm across most basins. We further note that we selected the 24h, 96h, and 120h lead times to study based on \Cref{fig:leadtime_breakdown} and conventions - where we observe that the 24h lead time represents the point at which models score similarly to each other (both deterministically and probabilistically), the 96h lead time presents the point at which the deterministic models all have an MSE close to or above 1 standard deviation, and 120h is the maximum lead time reported by some operational products (e.g., the 5-day NHC forecast) .
\par With regards to Claudia, we see that at 24h all three models are able to somewhat capture the behavior of the storm. At this lead time, the MLR provides a mostly too-intense prediction with a correspondingly high uncertainty that captures the observation in the 95\% confidence interval. The ANN mostly corrects the too-intense prediction made by the MLR, but relies on predicting a much larger uncertainty around the time of the peak (i.e., around 2020-01-14). At this lead time, the UNet model predicts a mean value that closely follows the observed track, though the prediction uncertainty is much larger compared to both other models.
\par At 96h, we see that the model whose mean behavior is best able to capture the observation is the MLR, though it once again appears to predict an overly smoothed, slightly more intense time series (except for the peak). Notably, at this lead time the predictions made by the MLR are much more uncertain compared to the predictions made at shorter lead times. At this lead time, we see a large deviation between the behaviors of the MLR and the ANN - where the mean predicted by the ANN is significantly lower than what was observed and where the intensity peaks in the observation fall outside the 95\% confidence interval of the prediction. At this lead time, the UNet model behavior is much more erratic, where the structure of the mean time series deviates significantly from the structure of the observation time series, and the observation lies mostly on the lower end of the predicted interval.
\par Finally, at 120h we see that the model that is best able to capture the structure of the observation is the MLR, though the uncertainty in the prediction is quite large. The ANN by comparison is unable to capture the peak within the limits of its prediction interval. Finally, the UNet prediction's behavior is quite erratic and is overconfident in both its over and under predictions.

\section{Conclusion}

While most NeWMs do not natively forecast wind gusts, we show that post-processing their meteorological fields allows for improved predictions of TC intensity compared to their unprocessed outputs. Post-processing models trained on NeWM outputs outperform those trained solely on ERA5 initial conditions, confirming the added value of the global atmospheric information encoded in NeWMs.

We introduced a tracking-independent post-processing setup and compared different post-processing algorithms. ANNs provided the best overall probabilistic skill, though case studies revealed that this does not always translate to realism at extended lead times. CNNs, particularly UNets, performed well at short lead times but degraded rapidly beyond 72-96h lead times. This challenge highlights the need for shared benchmarks to further engage the ML community in designing or post-processing NeWMs for extremes \citep{olivetti2024}.

Simple feature engineering, such as storm-centric masking, improved model performance. For example, masked linear models learned physically plausible patterns, such as increasing forecast uncertainty with lead time. Despite challenges at long lead times, our results show that even lightweight models can extract useful predictive signals from NeWM outputs. Their simplicity, combined with the low inference cost and global availability of deterministic NeWMs, supports the development of accessible end-to-end TC forecasting tools, complementing more computationally intensive generative or physics-based forecasting models.

Future work could explore larger training datasets and expanded predictor sets, including temporal features and variables used in operational statistical schemes \citep{NHC2022}. More specifically, future work could be expanded to consider oceanic data, given how important air-sea interactions have been in the intensification of storms \citep{nickerson2025rapid, zhang2023modulation, SrokaEmanual2021Review}. This is especially true considering that NeWMs have generally been trained as atmospheric models and generally do not predict ocean variables, or predict only the sea surface temperature \citep[e.g., as in][]{gencast24}. Thus, there are likely improvements to our framework derived from considering, e.g., sea subsurface temperature profiles, ocean heat content, surface enthalpy-fluxes, and air-sea mixing when predicting the intensity of tropical cyclones.
Our pipeline can be extended to forecast additional TC structural attributes recorded in IBTrACS \citep{chavas2022simple}, with the potential to reconstruct full wind fields via physics-guided methods \citep{eusebi2024realistic, Wang_2024}. Finally, the emergence of foundation models such as Aurora \citep{bodnar2025foundation} and generative NeWMs such as GenCast \citep{gencast24} opens new avenues for post-processing via low-rank adaptation or tail neural networks \citep{lehmann2025finetuningweatherfoundationmodel}, with potential generalizability gains that remain to be explored.

\tikzsetnextfilename{Figure_8}

\clearpage
\acknowledgments
We would like to thank Margot Sirdey at UNIL for their contributions to running experiments on HPC resources, Matthew Chantry at the ECMWF for their help with setting up our ai-models workflow, Yair Cohen at NVIDIA for their help with setting up our earth2mip workflow, and all of the members at the ECMWF and NVIDIA whose efforts made this research possible. We are especially grateful to Monika Feldmann, whose guidance was instrumental to this project, and Jordi Bolibar whose observations improved this work.
Tom Beucler further acknowledges support from the Swiss National Science Foundation (SNSF) under Grant No. 10001754 (``RobustSR'' project).

%
%
\datastatement
The code we used in the development of our study can be found at \url{https://doi.org/10.5281/zenodo.16893255}, and a singularity container with the virtual environment required to run the code is provided at \url{https://doi.org/10.5281/zenodo.16893255}. PanguWeather and FourCastNetv2 were run using the ECMWF's AI-models library, made available at \url{https://github.com/ecmwf-lab/ai-models/}. 

%






%



\FloatBarrier
\bibliographystyle{ametsocV6}
\bibliography{references,Louis_Thesis_bibliography}


\section*{Supplementary material}

# Supplementary Material: Global Forecasting of Tropical Cyclone Intensity Using Neural Weather Models

Milton Gomez*[a,b], Louis Poulain--Auzeau[a,b], Alexis Berne[c], and Tom Beucler[a,b]

[a]*Faculty of Geosciences and Environment, University of Lausanne, Lausanne, VD, Switzerland*
[b]*Expertise Center for Climate Extremes, University of Lausanne, Lausanne, VD, Switzerland*
[c]*Environmental Remote Sensing Laboratory, EPFL, Lausanne, VD, Switzerland*

This document supplements *Global Forecasting of Tropical Cyclone Intensity Using Neural Weather Models*, and includes three sections:

- Section S-1 - Supplemental Graphs. Here we provide feature distribution histograms for masked and unmasked inputs, as well as additional shapley value plots.

- Section S-2 - Supplemental Diagrams. Here we provide the architecture diagrams for the CNNs used in our study.

- Section S-3 - Regularization and Feature Selection Strategies. Here we provide additional information on the regularization and feature selection strategies in conducted experiments.

- Section S-4 - Supplemental Case Studies. Here we provide additional case study graphs.

*Corresponding Author*: milton.gomez@unil.ch

# S-1  Supplemental Graphs

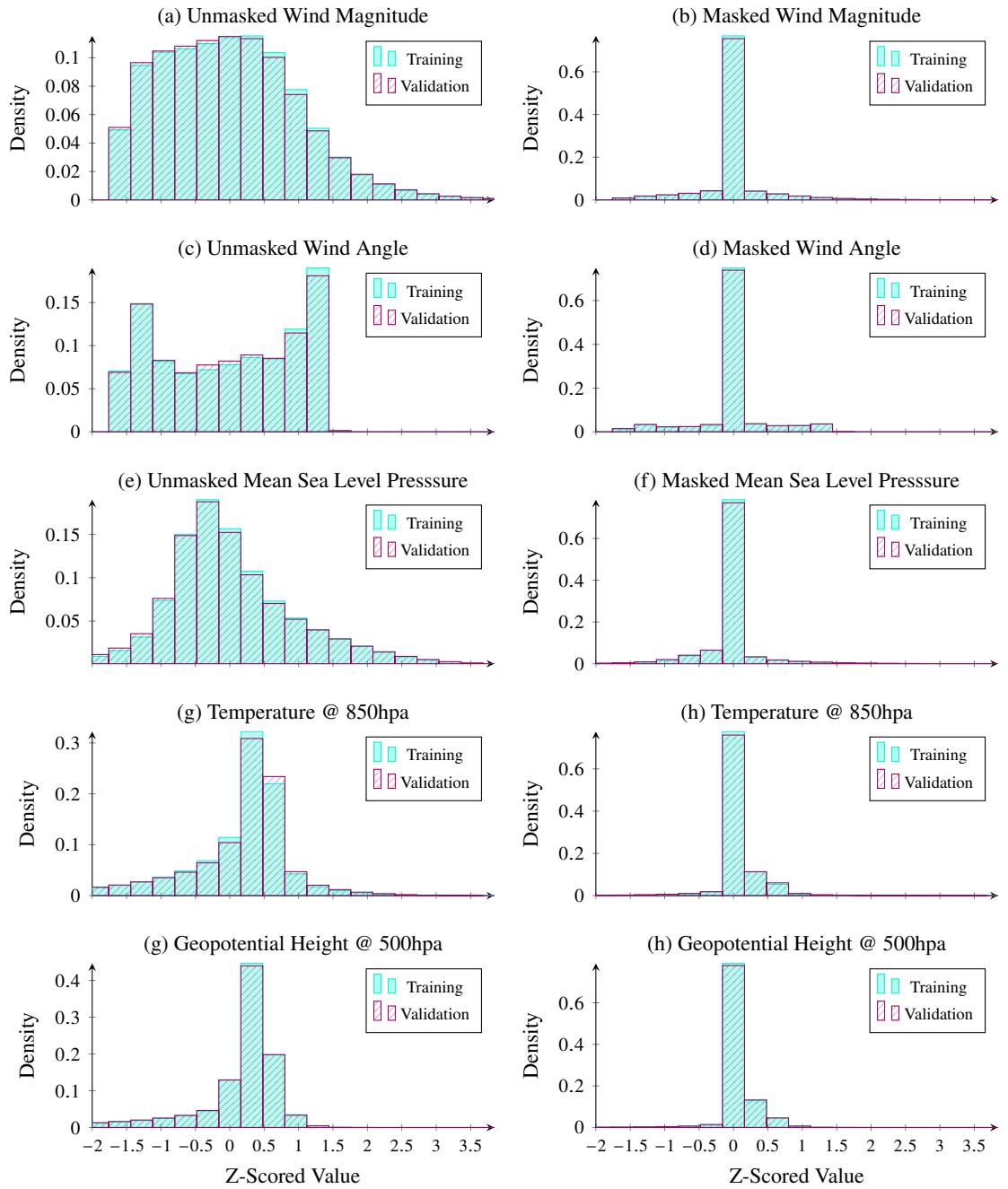

Figure S-1: Input Feature Distributions for the training and validation sets

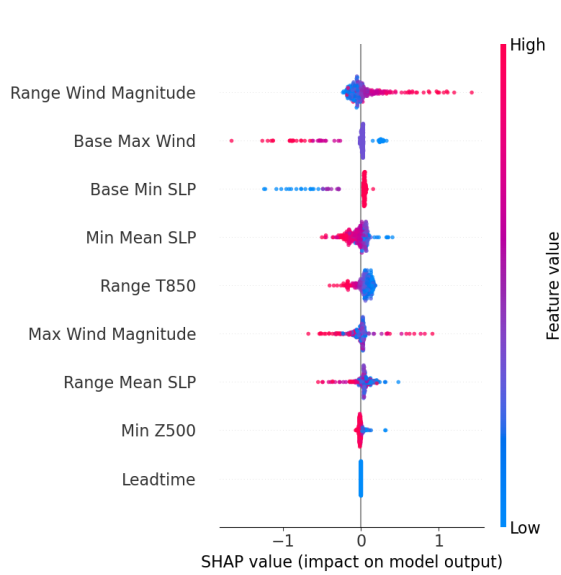

(a) Shapley values for the predicted mean ($\mu$) by the ANN

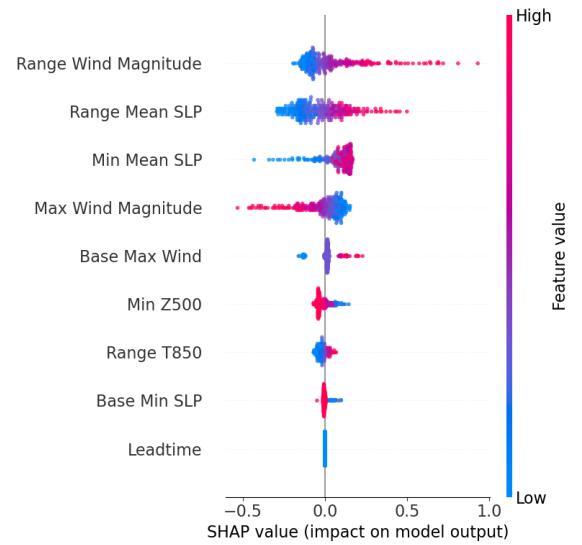

(b) Shapley values for the predicted standard deviation ($\sigma$) by the ANN

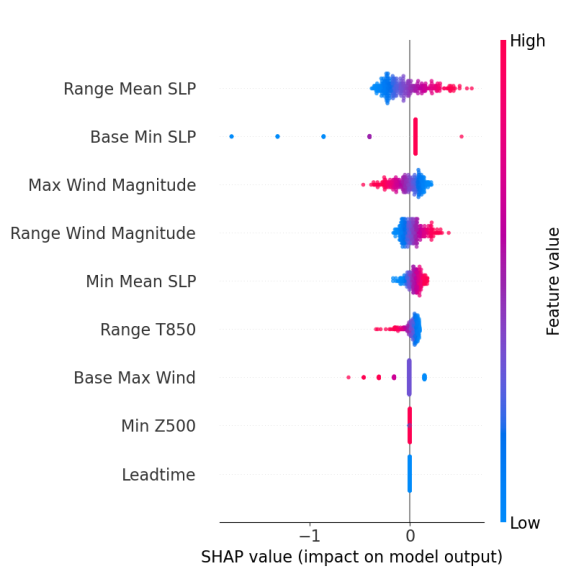

(c) Shapley values for the predicted mean ($\mu$) by the MLR

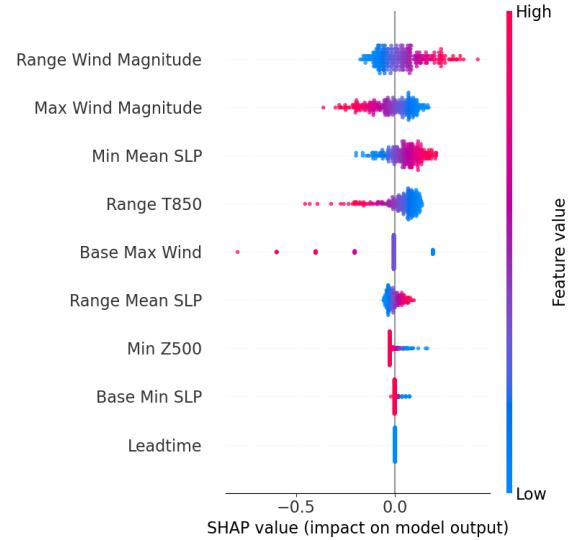

(d) Shapley values for the predicted standard deviation ($\sigma$) by the MLR

Figure S-2: Shapley values explaining individual feature contributions to the predicted value, corresponding to probabilistic models trained on masked inputs. Figures corresponding to $\tau = 24$h

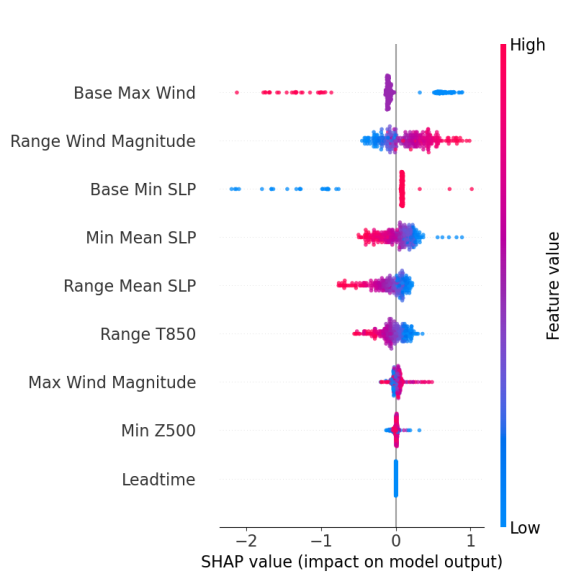

(a) Shapley values for the predicted mean ($\mu$) by the ANN

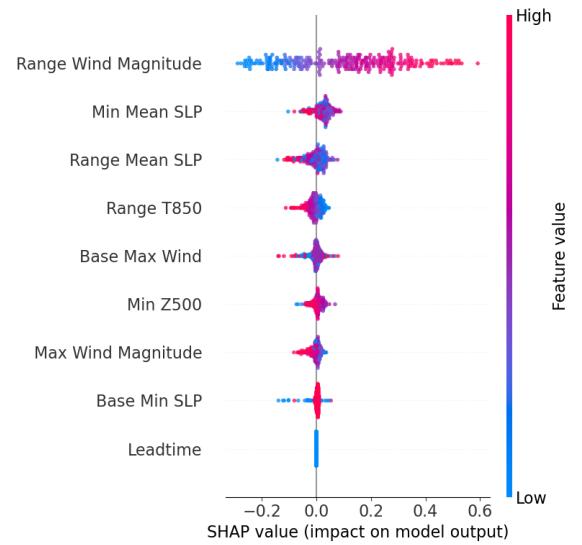

(b) Shapley values for the predicted standard deviation ($\sigma$) by the ANN

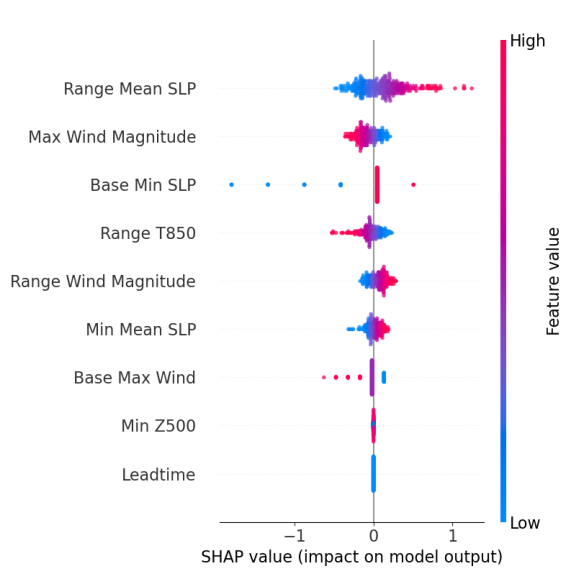

(c) Shapley values for the predicted mean ($\mu$) by the MLR

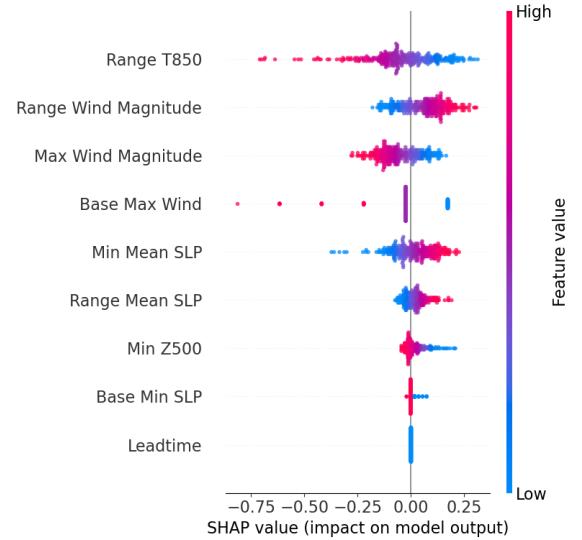

(d) Shapley values for the predicted standard deviation ($\sigma$) by the MLR

Figure S-3: Shapley values explaining individual feature contributions to the predicted value, corresponding to probabilistic models trained on masked inputs. Figures corresponding to $\tau = 72$h

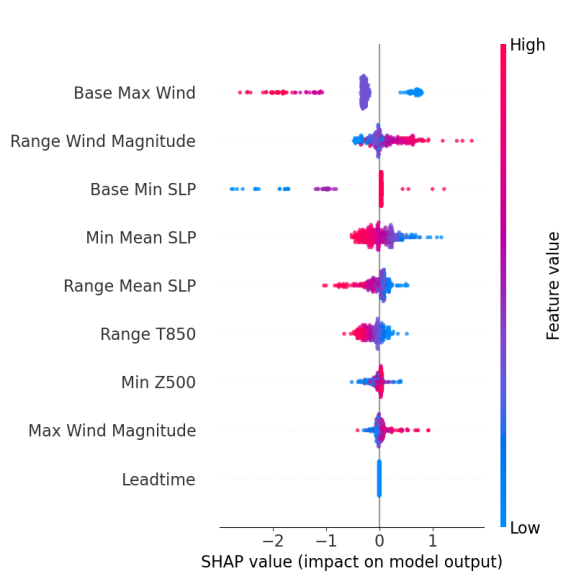

(a) Shapley values for the predicted mean ($\mu$) by the ANN

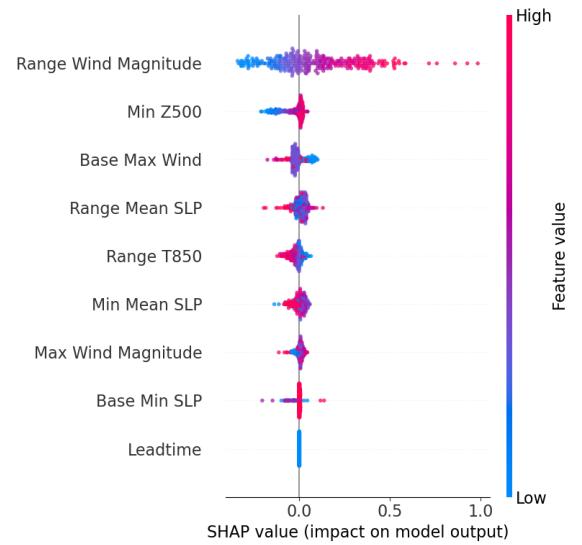

(b) Shapley values for the predicted standard deviation ($\sigma$) by the ANN

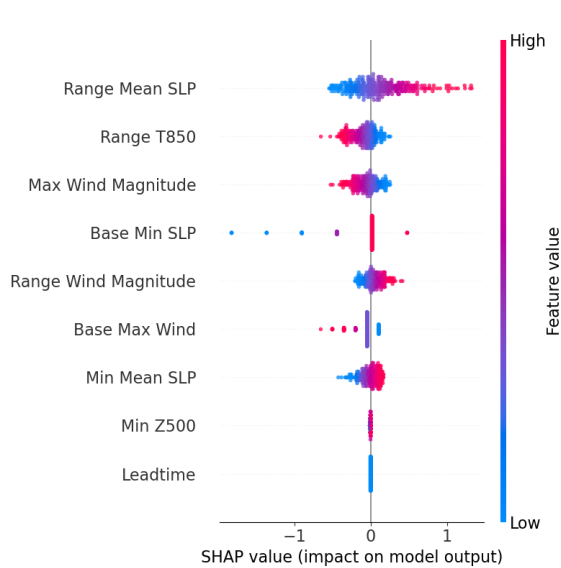

(c) Shapley values for the predicted mean ($\mu$) by the MLR

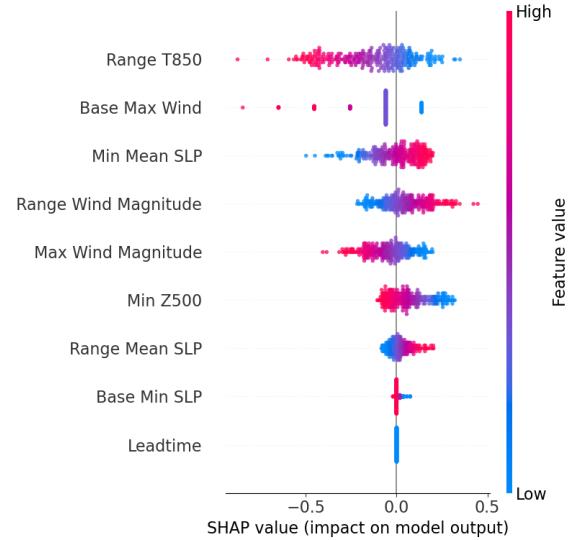

(d) Shapley values for the predicted standard deviation ($\sigma$) by the MLR

Figure S-4: Shapley values explaining individual feature contributions to the predicted value, corresponding to probabilistic models trained on masked inputs. Figures corresponding to $\tau = 120h$

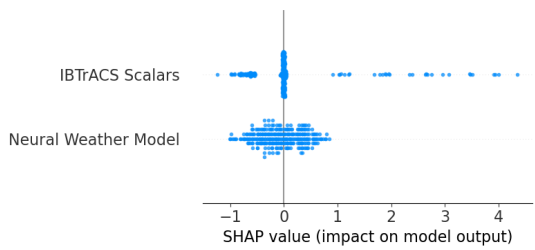

(a) Shapley values for the predicted mean ($\mu$) by the ANN

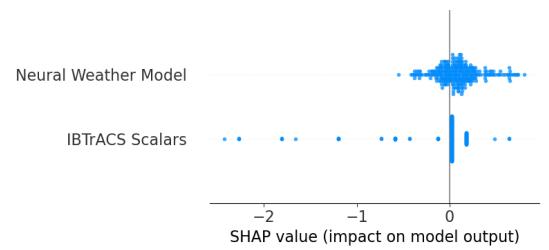

(b) Shapley values for the predicted mean ($\mu$) by the MLR

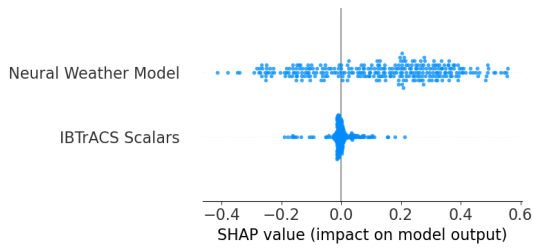

(c) Shapley values for the predicted standard deviation ($\sigma$) by the ANN

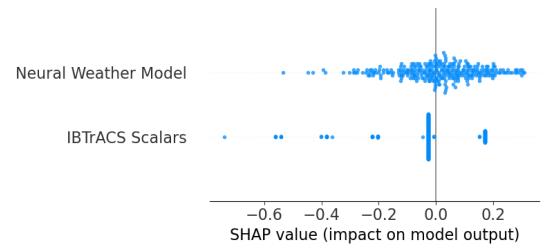

(d) Shapley values for the predicted standard deviation ($\sigma$) by the MLR

Figure S-5: Shapley values explaining the contribution by IBTrACS scalars vs neural weather model outputs to the predicted value, corresponding to probabilistic models trained on masked inputs. $\tau = 72h$

## S-2 Supplemental Diagrams

Here we present additional details on the CNN architectures used in our studies. When batch normalizaion was used, batch normalization layers were added between convolutional and pooling layers. Unless otherwise specified, any values take the default values provided by PyTorch. In the diagram, K represents the kernel size, S the step size, P the padding size, U the number of neurons/units, and the values within dropout layers show the probability of zeroing.

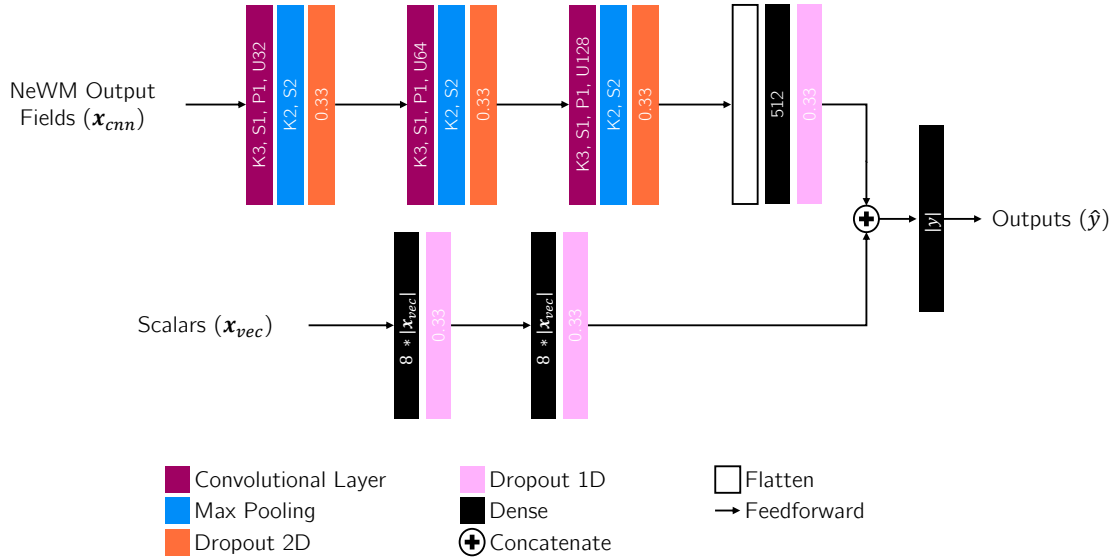

Figure S-6: CNN architecture

## S-3 Regulatization & Feature Selection Strategies

Our initial experiments with CNNs and all subsequent experiments with them resulted in the models overfitting to the training set. In order to attempt to address this, we relied on a number of strategies, including:

- L2 regulatization

- Dropout Layers

- Auxilliary Loss (using additional output units used to predict the training loss)

- Feature Ablation

- Setting the target to direct intensity values ($P_{min}$ , $|V_{max}|$) instead of intensification ($\Delta P_{min}$ , $\Delta |V_{max}|$)

- Training one model per leadtime; training a model for short term predictions and a second model for longer leadtimes; training a model that works across all lead times.

For CNNs we tested dropout rates by training a series of models to test at what point, if any, regularization led to similar behavior between the training and validation sets. The results are shown in fig. S-8, where we saw that regularization only degraded performance and did not mitigate overfitting without severely affecting performance.

Due to computational requirements, a much more experiment was run for the UNet architecture, and no new insights were derived from it.

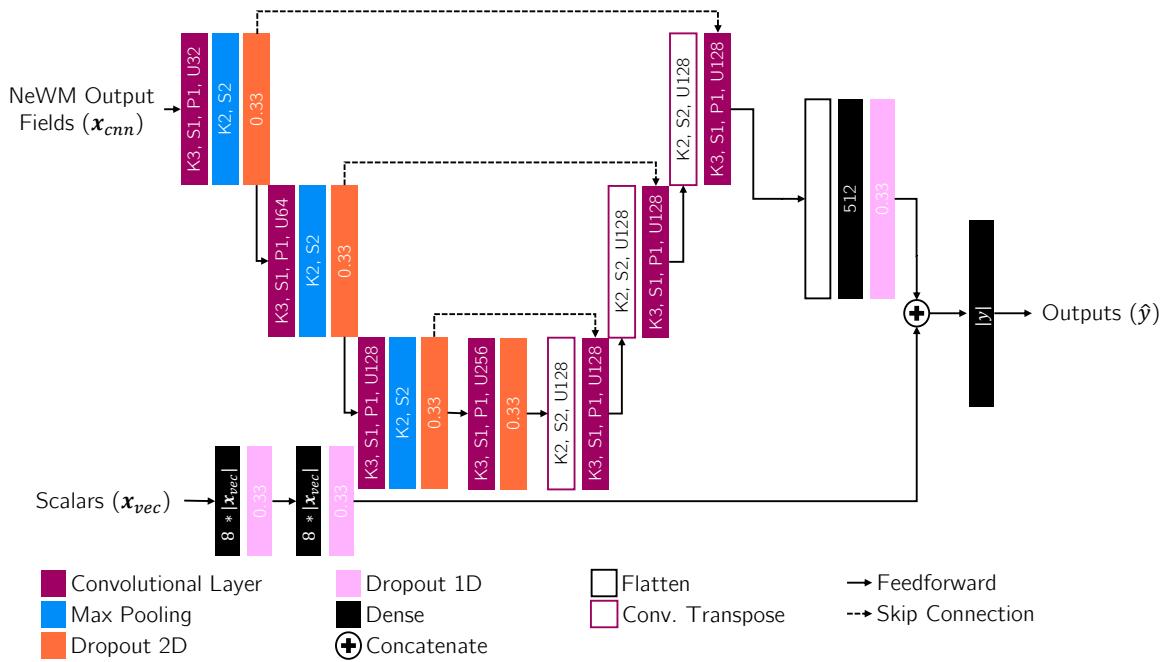

Figure S-7: UNet Architecture

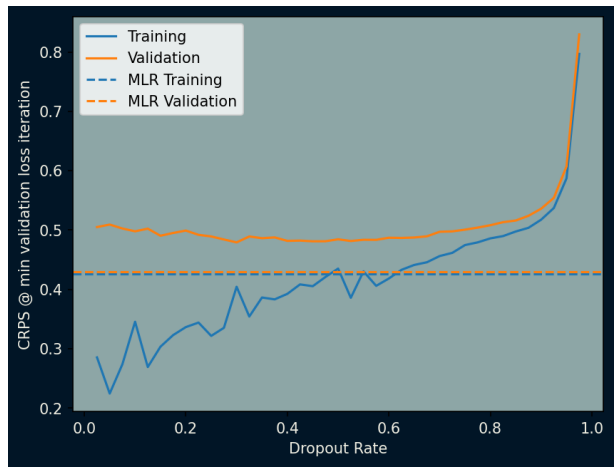

Figure S-8: Loss at best performing validation epoch vs dropout rate for a CNN architecture


\FloatBarrier

\end{document}